\documentclass[pdflatex,sn-mathphys-num]{sn-jnl}

\usepackage{graphicx}%
\usepackage{multirow}%
\usepackage{amsmath,amssymb,amsfonts}%
\usepackage{amsthm}%
\usepackage{mathrsfs}%
\usepackage[title]{appendix}%
\usepackage{xcolor}%
\usepackage{textcomp}%
\usepackage{manyfoot}%
\usepackage{booktabs}%
\usepackage{algorithm}%
\usepackage{algorithmicx}%
\usepackage{algpseudocode}%
\usepackage{listings}%
\usepackage{comment}

\usepackage{etoolbox}
\usepackage{lineno,hyperref}
\usepackage{mathtools}
\usepackage{pbox}
\usepackage{setspace}
\usepackage{float}
\usepackage{bbm}
\usepackage{bm}
\usepackage{subfigure}
\usepackage{xcolor, soul}
\usepackage{empheq}
\usepackage{pstricks}
\usepackage{color}

\newcommand*{\bmsf}[1]{\bm{\mathsf{#1}}}
\newcommand{\mbs}[1]{\mathbf{#1}}

\newcommand{\bZ}{{\bmsf{0}}}
\newcommand{\bA}{{\mbs{A}}}
\newcommand{\bB}{{\mbs{B}}}
\newcommand{\bC}{{\mbs{C}}}
\newcommand{\bD}{{\mbs{D}}}
\newcommand{\bF}{{\mbs{F}}}
\newcommand{\bI}{{\mbs{I}}}
\newcommand{\bH}{{\mbs{H}}}
\newcommand{\bK}{{\mbs{K}}}

\newcommand{\bT}{{\mbs{T}}}
\newcommand{\bX}{{\mbs{X}}}
\newcommand{\ba}{{\mbs{a}}}
\newcommand{\bn}{{\mbs{n}}}
\newcommand{\bu}{{\mbs{u}}}
\newcommand{\bx}{{\mbs{x}}}

\newcommand{\sig}{{\mbg{\sigma}}}

\newcommand{\PIso}{{\Psi}_{\rm iso}}
\newcommand{\PVol}{\Psi_{\rm vol}}
\newcommand{\Ip}{\bar{I}_1}
\newcommand{\Is}{\bar{I}_2}

\newcommand{\Cm}{\overline{\bC}}
\newcommand{\Fm}{\overline{\bF}}

\newcommand{\PV}{\Psi_{\rm vol}}
\newcommand{\PI}{\Psi_{\rm iso}}

\newcommand{\PA}{\Psi_{\rm aniso}}
\newcommand{\tr}{{\rm tr}}

\def\PV{{\Psi_{\rm vol}}}
\def\PI{{\Psi_{\rm iso}}}
\def\PA{{\Psi_{\rm aniso}}}

\def\Co{{\overline \bC}}
\def\Iu{{\overline I_1}}

\def\Iqs{{\overline I^*_4}}
\def\sig{{\sigma^2_{I_4}}}

\usepackage{cleveref}
\crefname{equation}{Eq}{Eq's}

\renewcommand{\figurename}{Figure}

\theoremstyle{thmstyleone}%

\theoremstyle{thmstyletwo}%

\theoremstyle{thmstylethree}%

\graphicspath {{./}{./Figures/}}

\begin{document}

\title[MicroStructured Cornea]{A coupled multiscale model of the human cornea accounting for the collagenous microstructure and the extracellular matrix}


\author{\fnm{Christopher} \sur{Miller}}
\equalcont{These authors contributed equally to this work.}

\author[1]{\fnm{Maria Laura} \sur{De Bellis}}
\equalcont{These authors contributed equally to this work.}

\author*[2]{\fnm{Anna} \sur{Pandolfi}}\email{anna.pandolfi@polimi.it}
\equalcont{These authors contributed equally to this work.}

\affil[1]{\orgdiv{Department of Engineering and Geology}, \orgname{University of Chieti-Pescara}, \orgaddress{\street{Viale Pindaro 42}, \city{Pescara}, \postcode{10587}, \country{Italy}}}

\affil*[2]{\orgdiv{Department of Civil and Environmental Engineering}, \orgname{Politecnico di Milano}, \orgaddress{\street{Piazza Leonardo da Vinci 32}, \city{Milano}, \postcode{20133}, \country{Italy}}}

\abstract{We present a micro-structurally based finite element model of the human cornea, where we explicitly describe the collagen-crosslink architecture in terms of a trusswork of non-linear struts, and the extracellular proteoglycan matrix in terms of continuum solid elements. We regard the cornea as a composite material with strongly non-linear properties within a finite kinematics framework. This innovative approach is based on two previously developed models, each of which has some drawbacks in describing stromal tissue degeneration. Separation of the continuum phase from the collagen microstructure allows a more realistic capture of the macroscopic phenomena observed in keratoconus pathologies, such as localized deformation.}

\maketitle

\section{Introduction}\label{sec1}

\begin{figure}[!h]
    \centering
    \includegraphics[width=0.7\textwidth]{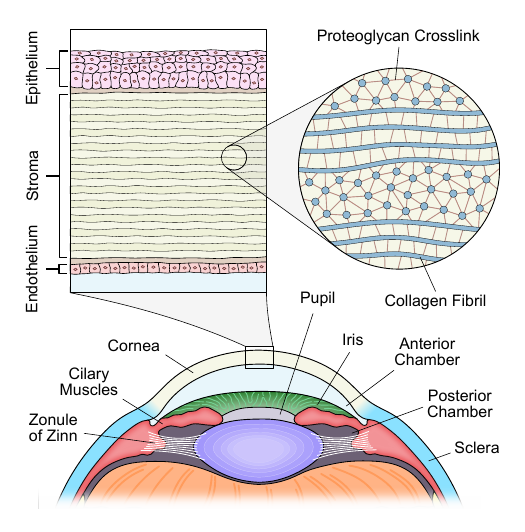}
    \caption{A schematic diagram of the anterior segment of the eye, displaying the repeated structure of the cornea. A magnified view of the stroma demonstrates the organisation of the collagenous microstructure into clearly defined lamellae consisting of differentially aligned collagen fibrils interconnected by proteoglycan crosslinks (not to scale).}
    \label{fig:fig1}
\end{figure}

The cornea is a quasi-spherical transparent shell occupying the outermost part of the eye, that covers the iris, pupil, and anterior chamber, see Fig.~\ref{fig:fig1}. Whilst it serves to protect the eye from infiltrates and ultraviolet radiation, the cornea's primary function is the refraction of light entering the eyeball. It is responsible for approximately two-thirds (43 diopters) of the eye's total optical power \cite{mishima:1968}. The cornea is made up of several distinct layers through its thickness, each fulfilling a specific role toward the overall functioning of the eye. However, it is the central stroma, making up 90\% of the total volume, that dominates the biomechanical properties of the cornea. The stroma consists of a complex architecture of extracellular matrix (ECM) proteins whose specialised structure confers considerable tensile stiffness to the cornea. The tissue is hierarchically organised into numerous ($\approx$300-500) lamellae running parallel to the corneal surface, each of which is composed of aligned bundles of collagen fibrils formed into ribbon-like sheets, with adjacent lamellae arranged at differing angles \cite{bron:2001}. Throughout the collagenous microstructure, fibrils are interconnected by proteoglycan (PG) crosslinks which facilitate effective load-transfer and modulate fibril spacing through a careful balance of attractive-repulsive forces \cite{maurice:1957}. 

The analysis of the corneal microstructure via X-ray imaging has demonstrated that more than 60\% of collagen fibrils have a randomly distributed orientation over the planes tangent to the mid-surface of the cornea, with the remaining 40\% exhibiting a more characteristically orthogonal arrangement \cite{abahussin:2009}. In the central cornea, collagen is clearly oriented in the nasal-temporal (NT) and superior-inferior (SI) directions, as a natural response to the activity of the eyelids and eye muscles \cite{meek:1987}. Additionally, a prominent circumferential-radial alignment is seen at the periphery of the cornea where it connects with the sclera \cite{aghamohammadzadeh:2004}. The local orthogonality of collagen fibrils is integral to the cornea's ability to sustain a mechanically-stable spherical shape under the action of an intraocular pressure (IOP) of $\approx$ 16 mmHg and, critically, ensures the optimal deviation of the light rays onto the retina.

Clinical observations suggest that alterations to the cornea's collagenous architecture are linked to irreversible changes in its spherical geometry, disrupting the proper passage and refraction of light into the eye. A prominent example is keratoconus, a progressive non-inflammatory disorder where a loss in fibril organisation leads to the cornea assuming a conical shape and thinning locally \cite{rabinowitz:1998}. This maladaptation can impair a patient's vision considerably, causing symptoms such as irregular astigmatism and high myopia. Additionally, abrupt changes in curvature may result in endothelial rupture which can severely impact the regulation of the stroma's chemical composition. The etiology of keratoconus and the underlying processes driving its advancement are not fully understood, but its occurrence has been attributed to a range of biological, chemical, genetic, mechanical, and environmental factors \cite{rabinowitz:1998}.

Given the many challenges that accompany the experimental assessment of the corneal microstructure, the identification of the deformation mechanisms underpinning healthy functionality and the changes corresponding to a diseased or damaged state are particularly troublesome \cite{simonini:2022}. Consequently, this has led to the pursuance of computational modelling methods aimed at reproducing the in vivo loading circumstances and replicating the mechanical behaviour of the cornea. Beyond their use as a tool to bolster our comprehension of corneal mechanics, they hold great potential toward a host of clinical applications, either in the design and optimisation of effective treatment strategies or through their direct introduction within the clinical workflow. Examples include the simulation of contact/contactless testing \cite{simonini:2016b}, the modelling of refractive surgeries (LASIK, SMILE, etc.) \cite{sanchez:2014} and surgical tools (Keratome, laser, etc.), as well as predicting the progression of ectatic disorders such as keratoconus \cite{pandolfi:2006}.

The majority of the numerical studies that have investigated the biomechanical response of the cornea have taken a continuum-based approach. Models of this kind have been employed extensively toward the modelling of numerous soft tissue types, as they can capture fundamental properties such as nonlinearity and anisotropy by accounting for tissue microstructure through a set of reinforcing fibers embedded within a continuum material, as well as other features such as fiber dispersion and spatially varying material stiffness \cite{pandolfi:2012,miller:2021,miller:2022}. Generally speaking, whilst it is clear that continuum material descriptions of soft biological tissues can include the influence of mechanically significant proteins and deliver physiologically reasonable outcomes, they often fail to address how specific ECM components relate to each other on a structural basis. They provide a homogenised mechanical response and as such, do not readily integrate information concerning the positioning and combined organisation of different constituents \cite{pandolfi:2019}. 

In the context of corneal modelling, continuum models are unable to suitably describe the complex interlinking of collagen fibrils via PG crosslinks, and how their functional interdependency gives rise to macroscopic tissue behaviour. The characterisation of collagen and crosslink mechanical properties is especially important in the case of models aimed at simulating the evolution of keratoconus, given the association of the pathology with degenerative changes to the collagenous microstructure. Continuum methodologies that have modelled conus formation through tissue weakening resulting from a region-specific temporal reduction of material stiffness have only been able to describe a moderate change in the cornea's spherical shape with limited localised thinning \cite{pandolfi:2006,simonini:2022}. Models of this kind provide unsatisfactory results as they cannot adequately reflect the loss of structural integrity of the collagen skeleton. 

These limitations have prompted the development of microstructurally-motivated discrete models in an effort to provide a more authentic portrayal of the cornea's architecture at sub-macro length scales, whereby, the main structural components of the stroma, collagen and PG crosslinks, are explicitly incorporated within a three-dimensional network of structural trusses \cite{pandolfi:2019}. The first deployment of this framework modelled all elements as linear elastic and by simply manipulating the spatial allocation of the stiffness, produced realistic deformation profiles for the healthy and diseased cornea \cite{pandolfi:2019}. Subsequent versions examined more histologically appropriate constitutive relations, such as a more realistic pseudo-chemical Lennard-Jones potential to describe crosslink behaviour \cite{pandolfi:2023} and a collagen description encompassing the stochastic variation of the orientational dispersion of fibrils \cite{debellis:2023}, with similarly successful outcomes. Other adaptations of the framework have introduced a damage-like scalar field whose evolution is governed by a reaction-diffusion equation \cite{pandolfi:2024}. This approach enables a spatial deterioration of element stiffness that can qualitatively predict conus formation.

The collective finding of the aforementioned studies has been the extent to which the weakening of chemical bonds between adjacent lamella heightens transversal shearing and compromises organ-level tissue stability, a feature consistent with experimental observations \cite{rabinowitz:1998}. The increased deformability of the system leads to localised bulging, pronounced thinning and the reshaping of the cornea, providing a better geometric approximation of keratoconus compared to previous attempts. By directly incorporating key aspects of the microstructure and their impaired mechanical functionality, the various iterations of the outlined discrete modelling framework have demonstrated a substantial improvement in our ability to describe the development of keratoconus.

However, a notable shortcoming of this modelling strategy is that it only considers the reinforcing architecture of the cornea, i.e., it consists solely of struts representing fibrous collagen and associated crosslinks. Neglecting the ECM in which the collagenous microstructure is embedded renders this an incomplete and unrealistic representation of corneal tissue. Going forward, if the true potential of this simple yet innovative approach is to be fully realised, the presence of the remaining non-collagenous isotropic ECM needs to be appropriately accounted for. 

With this in mind, the present work proposes the superposition of the existing discrete microstructural framework with a continuous representation of the non-collagenous matrix, which is modelled as a classical hyperelastic Mooney-Rivlin material. Trusses representing generalised collagen fibrils and PG crosslinks are described by hyperelastic and Lennard-Jones models, respectively. Additionally, a {\it new} finite element truss formulation is presented, which correctly models the tissue microstructure by accounting for material nonlinearity and finite kinematics. The coupled multiscale model is calibrated to provide mechanical behaviour representative of the human cornea. The influence of the geometric discretisation is investigated, and the ability of the model to characterise keratoconus is compared with previously reported computational models.

The organisation of the manuscript is as follows. In Section~(\ref{sec2}), we introduce the geometry of the cornea, the constitutive models based on microstructural concepts, and the novel aspects of the finite element implementation. Numerical results are collected in Section~(\ref{sec3}), and a discussion of the proposed model, identifying its significance and potential applications, is reported in Section~(\ref{sec4}).

\section{Materials and methods}\label{sec2}

\subsection{Corneal geometry and microstructure}\label{sec2.1}

To arrive at clinically meaningful results, modern advanced computational models of the cornea are built directly from clinical data, whereby patient-specific geometries are constructed through the application of a sophisticated interpolation procedure to a cloud of surface points obtained from corneal topography images \cite{simonini:2015, montanino:2023}. In general, coordinates are in reference to two axes that lie on the vertical plane orthogonal to the optic axis, $x$ corresponding to the NT direction, and $y$ corresponding to the SI direction, with the optic axis taken as the $z$ axis. In the subsequent discussion, we refer to the NT-SI plane as the corneal plane. In this study, the chosen geometry is characterised by a set of shape-related parameters commonly referred to in a clinical setting, which are listed in Table~\ref{Tab1}.
\begin{table}[!h]
\caption{Geometric parameters relating to the anterior and posterior surfaces of the healthy cornea, described as two biconic surfaces. The reference plane contains the $x$ (NT) and $y$ (SI) axes of the model.}
\label{Tab1}
\begin{tabular}{l r c}
\hline
\noalign{\vskip 0.15cm}
\hspace{0.2cm} & \textbf{Value} & \textbf{Unit}\cr
\noalign{\vskip 0.1cm}
\hline
\noalign{\vskip 0.2cm}
\hspace{0.2cm}\textbf{General}\cr
\hspace{0.6cm}{Central thickness}&{0.57}&\rm{mm}\cr
\hspace{0.6cm}{Apex elevation}&{2.48}&\rm{mm}\cr
\hspace{0.6cm}{In-plane diameter}&{10.60}&\rm{mm}\cr
\noalign{\vskip 0.1cm}
\hspace{0.2cm}\textbf{In-plane orientation}\cr
\hspace{0.6cm}{Steepest meridian NT}&{0}&\rm{deg}\cr
\hspace{0.6cm}{Flattest meridian SI}&{90}&\rm{deg}\cr
\noalign{\vskip 0.1cm}
\hspace{0.2cm}\textbf{Anterior surface}\cr
\hspace{0.6cm}{Steepest meridian radius}&{7.56}&\rm{mm}\cr
\hspace{0.6cm}{Flattest meridian radius}&{7.41}&\rm{mm}\cr
\hspace{0.6cm}{Asphericity coefficients}&{1.50}&\rm{mm}\cr
\noalign{\vskip 0.1cm}
\hspace{0.2cm}\textbf{Posterior surface}\cr
\hspace{0.6cm}{Steepest meridian radius}&{6.47}&\rm{mm}\cr
\hspace{0.6cm}{Flattest meridian radius}&{6.07}&\rm{mm}\cr
\hspace{0.6cm}{Asphericity coefficients}&{1.00}&\rm{mm}\cr
\noalign{\vskip 0.2cm}
\hline
\end{tabular}
\end{table}

\begin{figure}[!h]
    \centering
    \includegraphics[width=1.0\textwidth]{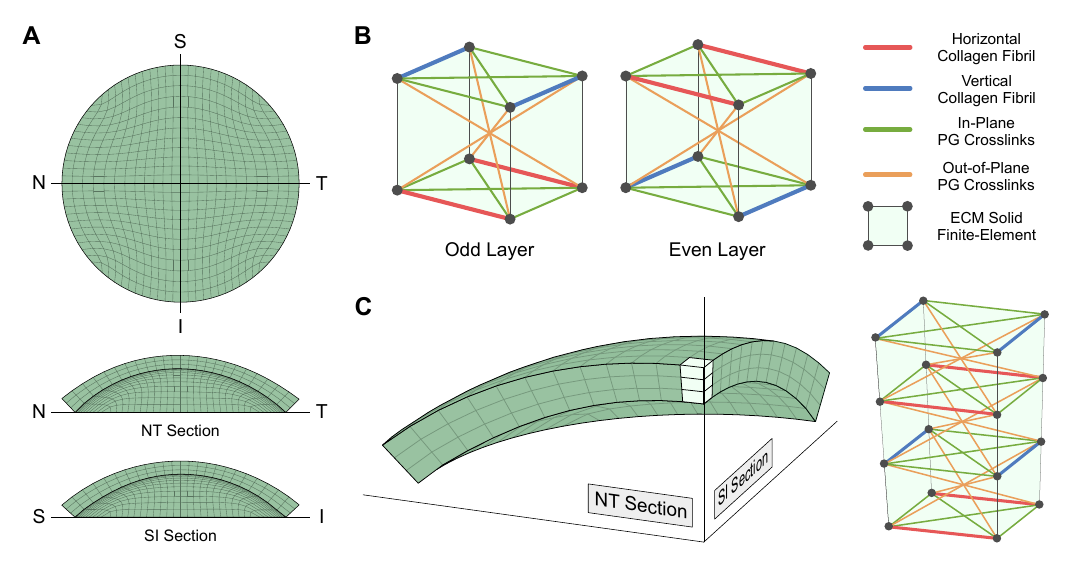}
    \caption{(A) An example of a finite element discretisation of the human cornea, consisting of 2,500 nodes and 1,728 8-noded hexahedral elements. (B) The configuration of the unit cell characterising the generalised microstructure of the cornea, with components describing collagen fibrils, PG crosslinks, and the remaining ECM. The multi-layer assembly of unit cells requires the distinction between those belonging to odd/even layers, so that the ideal organisation of the stromal architecture is replicated. (C) An assembly of multiple unit cells across the thickness of the corneal geometry at its apex, demonstrating how, over the entire geometry, this will lead to quasi-parallel surfaces of aligned collagen fibrils, ({\it lamina}), sequentially alternating in direction.}
    \label{fig:fig2}
\end{figure}

The geometry is discretised into hexahedral finite elements using an in-house developed 3D grid-generating software, where the discretisation process is controlled by two parameters, i.e., the number of elements $N_M$ along the principal meridian diameters (NT and SI), and the number of elements $N_L$ across the thickness. An example of a generated mesh is provided in Fig.~\ref{fig:fig2}A. The collagenous microstructure, assembled from a series of truss elements, is then superposed onto each solid element to form a unit cell, the configuration of which is detailed in Section (\ref{sec2.2}).

The loading of the discretised geometry is provided by imposed tractions (Neumann boundary conditions) associated with the intraocular pressure at the posterior surface of the cornea. It is also necessary to impose the displacements (Dirichlet boundary conditions) at the external annulus of the geometry, i.e., the limbus, to properly account for the rotational freedom induced by the relative softness of the neighbouring sclera and iris.

It should be noted that the measured geometry of the cornea relates to the tissue in its deformed state under the action of the IOP. As such, the entirely accurate modelling of the cornea using the finite element method necessitates the recovery of its unstressed configuration. Be that as it may, given the added complexity this would bring to the current study and the fact that our primary motivation is to demonstrate the efficacy and robustness of the presented coupled multiscale model, we simply treat the measured cornea as our referential geometry. This issue will, however, be addressed in future work through the integration of our previously developed inverse stress analysis scheme \cite{pandolfi:2006} within the modelling framework.

\subsection{Unit cell definition}\label{sec2.2}

The microstructure of the cornea is characterised by the spatial repetition of a unit cell, where each unit cell consists of a solid 8-noded hexahedral finite element relating to the non-collagenous ECM material, and a series of truss elements interconnecting the various the nodes of the solid element, representing either a collagen fibril or a PG crosslink. 

As illustrated in Fig.~\ref{fig:fig2}B, we consider a unit cell to have two possible configurations, corresponding to whether it is located within an odd-numbered or an even-numbered layer, where a layer is defined as a group of unit cells existing at the same depth through the thickness of the cornea. The first layer is on the posterior side of the cornea, whilst the number allocated to a layer increases successively in a posterior-to-anterior manner until the maximum is reached on the cornea's anterior side. An odd unit cell contains two vertical collagen trusses interspersed by two horizontal and two diagonal in-plane PG crosslinks located across its top surface, whereas the bottom surface contains two horizontal collagen trusses interspersed by two vertical and two diagonal in-plane PG crosslinks. Finally, four diagonal out-of-plane crosslinks link the nodes of the unit cells' top and bottom surfaces. An even unit cell is then simply the mirror of the odd unit cell in the z-axis. An example assemblage of several unit cells through the thickness of a three-layered discretised geometry is depicted in Fig.~\ref{fig:fig2}C. It can be seen that in those instances where multiple adjacent unit cells contain entirely coincident trusses, we model only one truss element that belongs to all neighbouring unit cells.

The configuration of unit cells in this way, when assembled to form the corneal geometry in its entirety, gives rise to quasi-parallel surfaces of aligned collagen fibril trusses separated equidistantly throughout the corneal thickness, which we will term laminae. The direction of alignment alternates sequentially in an approximately orthogonal fashion between the horizontal and vertical directions for each successive lamina. Therefore, in accordance with their hypothesised mechanical function, the in-plane crosslinks act to distance the collagen fibrils within a lamina, whilst the out-of-plane crosslinks effectively separate neighbouring laminae. From the arrangement of the PG crosslinks relative to the collagen fibrils, it is readily apparent that their presence will confer the generated trusswork with heightened structural stability compared to if they were absent.

In furtherance to the discussion of the geometric discretisation procedure in the previous section, the mesh is generated in such a way that ensures the orientation of collagen truss elements properly reflects the spatial variation in collagen fibril direction (averaged across the thickness) as observed during the experimental imaging of the cornea \cite{aghamohammadzadeh:2004}. Accordingly, the main orientation of the fibrils gradually varies from an orthogonal arrangement at the centre, which follows the NT and SI directions, to an orthogonal arrangement at the limbus, where fibrils run circumferentially and radially. Importantly, to ensure an equal number of collagen fibril lamina aligned in the horizontal and vertical directions, the framework is restricted to the consideration of discretised geometries with an odd total number of layers, $N_L$.

All microstructural-based approaches require a set of assumptions to be made regarding the length-scales reflected in their mathematical framework, and the presented coupled multiscale model is no different. As the wholly realistic portrayal of the tissues' natural micro/nano-scales in a structural sense would require an inordinate number of elements, rendering any simulation far too computationally demanding, the approach taken is to instead assemble a network of trusses representing a more coarse \textit{generalised} form of the collagenous architecture. The individual response of each truss element, in fact, corresponds to a collection of a given constituent. Modelling the hierarchical organisation of collagen and PG crosslinks in this way, whilst a simplification compared to the ideal case, still retains our ability to investigate the structure-functional relationship of fibrillar collagen and PG crosslinks and how it imbues corneal tissue with its macroscopic mechanical stiffness and contributes to diseases such as keratoconus.

\subsection{Constitutive descriptions}\label{sec2.3}

Here we detail the 1D constitutive descriptions of the generalised collagen fibrils and proteoglycan crosslinks, as well as the 3D continuum description accounting for the combined response of the remaining ECM constituents, which are not represented by the assembled trusswork.

\subsubsection{Microstructural truss descriptions}\label{sec2.3.1}

The outlined descriptions consider each truss to be a hyperelastic body undergoing large deformations and, thus, employ finite kinematical theory. As such, the relative deformation measure used is the stretch $\lambda$, i.e., the deformed length $l$ of the truss, divided by $L$, its undeformed referential length, where $\lambda$ is considered to be uniform over the length of the truss. Furthermore, in keeping with experimental observations of incompressibility concerning fibrous collagen and proteoglycans, we assume all trusses to be conserve their volume when deformed. 

Given that the traditional treatment of trusses is in a force versus displacement setting, by analogy and for convenience, the formulation is developed in the material
reference frame, and therefore, the relations
\begin{align}\label{SC}
    P(\lambda) = \dfrac{\partial\psi}{\partial \lambda}\, ,\quad \mathcal{A}(\lambda)=\dfrac{\partial P(\lambda)}{\partial \lambda}=\dfrac{\partial^2\psi}{\partial\lambda^2}\, ,
\end{align}
provide the scalar first Piola-Kirchhoff stress $P$, and the associated scalar 
stiffness $\mathcal{A}$, of a truss as a function of $\psi$, its potential energy. It is also worth noting that in the 1D case, $P$ and $\lambda$ represent an energetic conjugate pairing.

\paragraph{Collagen fibril}\label{sec2.3.1.2}
The material behaviour of a truss representing a generalised collagen fibril is governed by a phenomenological constitutive description, well-established in the literature \cite{holzapfel:2000}, which accounts for the nonlinear stiffening of soft tissues arising from the recruitment of undulated collagen fibrils. We begin by introducing a quadratic potential energy per unit volume acting along the length of the truss
\begin{align}\label{Coll_SEDF}
    {\psi}_{\rm coll}(\lambda) =
    \dfrac{k_{\rm{1}}}{2k_{\rm{2}}}\left[ \exp{ \left\{ k_2\left(\lambda-1\right)^2 \right\}} - 1\right]\, ,
\end{align}
where $k_{\rm{1}}$ is the elastic stiffness of the fibril per unit area and $k_{\rm{2}}$ is a dimensionless elastic rigidity parameter. The first Piola-Kirchhoff stress is then defined as
\begin{align}\label{Coll_Sig}
    P_{\rm coll}(\lambda)=k_1 \left(\lambda-1\right)\,\exp\left\{k_2\left(\lambda-1\right)^2\right\}\, ,
\end{align}
which in the absence of the exponential term would reduce to a standard Neo-Hookean description. The stiffness is then found to be 
\begin{align}\label{Coll_Stiff}
  \mathcal{A}_{\rm coll}(\lambda)=k_1\left[1 + 2k_2\left(\lambda-1\right)^2\right]\exp\left\{k_2\left(\lambda-1\right)^2\right\}\, .
\end{align}
which completes our 1D constitutive description for a generalised collagen fibril.

\paragraph{Proteoglycan crosslink}\label{sec2.3.1.1}

To characterise the mechanical response of the trusses representing the generalised PG crosslinks, we adopt a Lennard-Jones (LJ) potential. LJ models were initially conceived to describe the energy of two interacting objects as a function of the distance between them. They are able to capture the repulsive forces of particles (e.g. atoms, molecules) at close distances, attractive forces at moderate distances, and the decay of interacting forces at infinite distances. 

For this reason, they have also been used to represent the state of equilibrium existing in PG's and the role this plays in modulating the spacing of adjacent collagen fibrils \cite{pandolfi:2023}. Specifically, PG's are prevented from assuming a fully extended conformation, which results in forces that tend to move fibrils closer together. However, PGs are hydrophilic, and the increased water volume leads to forces that push fibrils apart. A careful balance is reached, which in turn gives rise to specific interfibrillar distances.

To phenomenologically describe this behaviour, we introduce a potential energy per unit volume of the form
\begin{align}\label{PG_SEDF}
    \psi_{\rm pg}(\lambda)=\varepsilon_{\rm{}}\lambda^{-a}\left(\lambda^{-a}-2\right)\, ,
\end{align}
where $\varepsilon$ is the minimum potential energy per unit volume of the crosslink, i.e., the energy in the undeformed state. The first Piola-Kirchhoff stress is then denoted by
\begin{align}\label{PG_Sig}
    P_{\rm pg}(\lambda)  = 2a\varepsilon_{\rm{}}\lambda^{-(a+1)}\left(1-\lambda^{-a}\right) \, ,
\end{align}
where it is evident that the parameter, ${\varepsilon}_{\rm{}}$, dictates the peak stress of the crosslinks response, whilst the non-dimensional parameter, $a$, controls the change in stress with increasing axial deformation. The expression
\begin{align}\label{PG_Stiff}
  \mathcal{A}_{\rm pg}(\lambda) = 2\varepsilon_{\rm{}} a\lambda^{-2(a+1)}\left[2(a+1)-\lambda^a(2+a)\right]\, ,
\end{align}
then details the corresponding stiffness and concludes the non-linear constitutive description for a generalised PG crosslink 

\subsubsection{Continuum extracellular matrix description}\label{sec2.3.2}

The continuum description of the isotropic ECM is determined according to the classical decoupled volumetric-deviatoric formulation of the strain energy density, thus ensuring that incompressibility, a focal feature of corneal tissue, is effectively enforced. The strain energy density is consequently given by
\begin{align}\label{SEDF}
    {\Psi}_{\rm{ECM}}=\PVol(J)+\PIso (\Ip,\Is)\, ,
\end{align}
where $\PVol$ denotes a purely volumetric contribution that acts as a penalty term to impose the incompressibility constraint, and $\Psi_{\rm{iso}}$ is a purely isochoric contribution. The volumetric strain-energy takes the operative form
\begin{align}\label{SEDF_vol}
    \PVol(J)=\dfrac{K}{4}(J^2-1-2{\rm{log}}J)\, ,
\end{align}
which depends on the Jacobian $J={\rm{det}}\bF$, where $\bF=\partial\bx/\partial\bX$ is the deformation gradient. The coefficient $K$ is related to the bulk modulous of the material, a suitably high value of which effectively preserves the volume at a Gauss-point \cite{simo:1991}.

An isotropic hyperelastic Mooney-Rivlin model \cite{rivlin:1951} is assumed for the isochoric part of the deformation, based on its previous successful application to the cornea in describing the ECM \cite{pandolfi:2006}. Accordingly, $\PIso$ is defined by the relation
\begin{align}\label{SEDF_ECM}
    \PIso(\Ip,\Is)=\dfrac{\mu_1}{2}(\Ip-3)+\dfrac{\mu_2}{2}(\Is-3)\, ,
\end{align}
with $\mu=\mu_1+\mu_2$ denoting the shear modulous of the material, $\Ip={\rm{tr}}(\Cm)$ and $\Is=[({\rm{tr}}(\Cm))^2-{\rm{tr}}(\Cm^2)]/2$, corresponding to the first and second invariants of the modified right Cauchy-Green deformation tensor, itself described by, $\Cm=\Fm^{\mathrm{T}}\Fm$, where $\Fm={J^{-1/3}}\bF$ is the modified deformation gradient.

\subsection{Finite element formulation}\label{sec2.4}

The aspects of the finite element formulation relevant to the present multiscale model of the cornea are briefly recalled. All simulations are carried out using a specifically designed in-house software coded in the programming language C.

\subsubsection{Truss elements}\label{sec2.4.1}


We begin by considering a truss element consisting of two nodes $\{\rm{a},\rm{b}\}$ whose referential and deformed coordinates are given by the two sets of vectors $\{\bX_{\rm{a}}, \bX_{\rm{b}}\}$ and $\{\bx_{\rm{a}}, \bx_{\rm{b}}\}$, respectively. Consequently, the relation $\bu_{(\bullet)}=\bx_{(\bullet)}-\bX_{(\bullet)}$ provides the displacement vector for a given node. 

Following on from the constitutive descriptions outlined in Section~(\ref{sec2.3.1}), the internal forces acting at each node of the truss are denoted by
\begin{align} \label{Truss1}
  \bT^{}_{\rm{a}} = - P(\lambda)A~\bn\, , \quad \bT^{}_{\rm{b}} = P(\lambda)A~\bn\, ,
\end{align}
where $A$ is the referential cross-sectional area, $\bn=\left(\bx_{\rm{b}}-\bx_{\rm{a}}\right)/l$ is a unit vector defining the deformed truss's three-dimensional direction in space, and $P(\lambda)$ is the first Piola-Kirchhoff stress of the truss.

The linearisation of the equilibrium equations for each node with respect to the two displacement vectors ultimately yields a set of linear equations for an element, which in matrix representation reads
\begin{align} \label{Truss2}
\left[\begin{array}{c}
  \bT^{}_{\rm{a}} \\[6pt]
  \bT^{}_{\rm{b}}
\end{array}\right]=
\left[\begin{array}{cc}
  \bK^{}_{\rm{aa}} & \bK^{}_{\rm{ab}} \\[6pt]
  \bK^{}_{\rm{ba}} & \bK^{}_{\rm{bb}}
\end{array}
\right]
\left[\begin{array}{c}
  \bu_{\rm{a}} \\[6pt]
  \bu_{\rm{b}}
\end{array}\right].
\end{align}
The individual contributions to the element tangent stiffness matrix relate the change in the forces at a node to the change in the current position of a particular node. Each contribution is defined according to the relations
\begin{align} \label{Truss3}
 \bK^{}_{\rm{aa}} = 
 \bK^{}_{\rm{bb}} =  
 \left[\alpha(\lambda)-\beta(\lambda)\right] \left(\bn\otimes \bn\right)+\beta(\lambda)\bI \, , \quad 
 \bK^{}_{\rm{ab}} =
 \bK^{}_{\rm{ba}} =
 -\bK^{}_{\rm{aa}}\,,
\end{align}
where $(\bn\otimes\bn)$ is a second-order unit structural tensor containing the trusses directional information and $\bI$ denotes the identity tensor. The two terms in Eq.~(\ref{Truss3}a) are analogous to the constitutive and geometric quantities arrived at when deriving the tangent stiffness matrix for a solid finite element. The deformation-dependent scalars, $\alpha(\lambda)$ and $\beta(\lambda)$, have the units, force per unit length, and are defined by the expressions 
\begin{align} \label{Truss4}
  \alpha(\lambda) =\dfrac{\mathcal{A}(\lambda)A}{L} \, , \quad  \beta(\lambda) =\dfrac{P(\lambda)A}{l}\,.
\end{align}
Their derivation, arising from the directional derivative of the internal truss forces in the material frame, is provided in Appendix~\ref{AppA}. 

As was detailed in Sec.~\ref{sec2.1}, all trusses are generalised and thus characterise a collection of a given structural protein. Specifically, the unit cell, the fundamental building block of the corneal model, represents the up-scaling of low-scale microscopic components to the macroscale \cite{koery:2024}. Thus, each truss element must account for a certain quantity of the tissue material, which reduces proportionally with the size of the discretisation. This can be facilitated in a straightforward way by modifying the referential cross-sectional area $A$ allocated to each element. The internal forces and the tangent stiffness matrix contribution of each truss element are assumed to be a function of both $N_M$, the number of unit cells along the principal meridian, and $N_L$, the number of unit cells across the thickness. Thus, the area of all trusses in the assembled meshwork is defined as
\begin{align} \label{Weight}
A=w_{M}w_{L}\overline{A}
\end{align}
where $w_{M}=1/N_M$ and $w_{L}$ are scalar weighting factors referring to the meridian and thickness discretisation, respectively. Note that the collagen and crosslink trusses located entirely (both nodes) on the anterior or posterior surfaces are allocated a weighting of $w_{L}=1/2N_L$, whilst the remaining majority of the trusses are allocated a weighting of $w_{L}=1/N_L$. As the cross-sectional area carries no explicit physiological meaning, we set the input parameter $\overline{A}=1$ for all trusses such that the mechanical properties of each truss are then fully encapsulated by their respective constitutive parameters.

\subsubsection{Solid elements and boundary conditions}\label{sec2.4.2}

To model the ECM continuum material, we used standard linear 8-noded hexahedral isoparametric elements, whose formulation is well-established in the literature and can be found in standard textbooks, e.~g., \cite{bonet:2008,belytschko:2014}. 

The IOP exerted by the aqueous humour is assumed to act uniformly over the posterior surface of the cornea and to act exclusively on the posterior facets of the solid elements, since the collagen/crosslinks trusses cannot be loaded transversally. The pressure is translated into equivalent nodal forces by using the energetic equivalence that arises from the weak form of the linear momentum balance used in the Galerkin approach. The IOP acts in the direction normal to the posterior surface, and as such, the external loading of the system is a function of the cornea's current deformation state and must be recomputed at every time step.

Concerning the displacement boundary conditions at the limbus, previous numerical studies have demonstrated that the dominant mechanical effect of surrounding tissues upon the cornea is that they limit the occurrence of bending moments \cite{pandolfi:2008}. We therefore enforce that the cross-section of the cornea at the limbus preserves orthogonality conditions with respect to the deformed mid-section of the shell, reducing the engagement of the tissue in terms of stored energy, an aspect that complies with the concept of energy minimization governing the behaviour of biological homeostasis.

\subsubsection{Solution methodology}\label{sec2.4.4}

The solution of the quasi-static non-linear problem of the pressurised cornea requires the introduction of a simulation time-frame, $t_n=t_{n-1}+\Delta t$, where $\Delta t$ is the incremental time step. The finite element spatial discretization of the linear momentum equation leads, following assembly, to a non-linear algebraic system of equations that, at the time $t_{n}$, can be written in the form
\begin{align} \label{balance}
\bmsf{R}_{n}(\bmsf{u}_{n})=\bmsf{T}_{n}(\bmsf{u}_{n})-\bmsf{F}_{n} \rightarrow \bZ \, ,
\end{align}
where $\bmsf{T}_{n}(\bmsf{u}_n)$ is the global internal forces array, $\bmsf{u}_{n}$ is the global displacement array, and $\bmsf{F}_n = \bmsf{F}_{n-1} + \Delta \bmsf{F}$ is the global external forces array, with the residual array $\bmsf{R}_{n}(\bmsf{u}_{n})$ becoming $\bZ$ only when equilibrium is satisfied. 

In this modelling framework two solution strategies are implemented, the first of which is the traditional Newton-Raphson method. Linearization of the residual array introduces the tangent stiffness matrix and accompanying set of equations, that read
\begin{align} \notag
\bmsf{K}^k_{n} = \frac{\partial\bmsf{R}^k_{n}}{\partial\bmsf{u}}\,,\quad\bmsf{u}^{k+1}_{n} = \bmsf{u}^k_{n} - \left[\bmsf{K}^k_{n} \right]^{-1} \bmsf{R}^k_{n}\,,  
\end{align}
and are solved iteratively at each time point $t_{n}$, until global equilibrium, i.e., Eq.~(\ref{balance}), is satisfied to a predefined tolerance.

However, given the definition of a unit cell, the proposed multiscale model requires a significant number of elements for even moderately discretised corneal geometries, and the resulting size of the global tangent stiffness matrix can become very large, rendering the simulation cumbersome and computationally expensive. Furthermore, this model was conceived with its future application towards the simulation of corneal degeneration in mind. For such problems, progressive tissue softening can potentially lead to accompanying, often severe, numerical issues when using the Newton-Raphson scheme. A more convenient approach is therefore to utilise the dynamic relaxation method \cite{oakley:1995}, where we instead construct a critically damped pseudo-dynamic problem described with a fictitious time $\tilde{t}$ (independent of the timescale of the simulation), defined by the expression 
\begin{align} \label{SM1} 
    \bmsf{R}_{n}(\bmsf{u}_{n})
    =
    \bmsf{M}_{\rm{fict}}\ddot{\bmsf{u}}_{n}(\tilde{t})+\bmsf{D}_{\rm{fict}}\dot{\bmsf{u}}_{n}(\tilde{t})
    +\bmsf{T}_{n}(\bmsf{u}_{n},\tilde{t})-\bmsf{F}_{n}
    \rightarrow
    \bZ
\, ,
\end{align}
and designed for the fastest possible convergence toward the steady-state solution. The internal variables, $\dot{\bmsf{u}}$ and $\ddot{\bmsf{u}}$ are the nodal velocity and acceleration arrays, and $\bmsf{M}_{\rm{fict}}$ and $\bmsf{D}_{\rm{fict}}$ are the fictitious diagonal mass and damping arrays. The pseudo-dynamic problem in Eq.~(\ref{SM1}) is numerically integrated with respect to $\tilde{t}$ until the steady-state solution is reached ($\ddot{\bmsf{u}}=\dot{\bmsf{u}}=\bZ$) and global equilibrium is satisfied to a predefined tolerance. 

The pseudo-time $\tilde{t}_k=\tilde{t}_{k-1}+\Delta\tilde{t}$ at each iteration of the dynamic relaxation method is defined by $\Delta \tilde{t}$, an arbitrary chosen pseudo-time step. Since the actual density of the material is not relevant in the static problem, for each solid element the pseudo-time step is used to define an ideal density $\rho_e$ such that the critically stable time step $\Delta t_e$, defined by the Courant–Friedrichs–Lewy dynamic stability condition, coincides with the chosen $\Delta \tilde{t}$. For a given element size $h_e$, the critical $\Delta t_e$ is defined as
\begin{align} \label{SM2}
    \Delta t_e = \frac{h_e}{c_e} \approx h_e \sqrt{\frac{\rho_e}{E_e}} = \Delta \tilde{t} \quad \rightarrow
    \quad
    \rho_e = E_e \frac{\Delta \tilde{t}^2}{h_e^2} \,,
\end{align}
where $c_e$ is the longitudinal wave speed of the material and $E_e$ is the {\it average} elastic modulous of the element. The element mass matrix is then obtained via the Galerkin approximation by using the element interpolation functions, row lumped to facilitate the numerical solution. 

The damping matrix $\bmsf{D}_{\rm{fict}}$ is assumed to scale linearly with $\bmsf{M}_{\rm{fict}}$ by a weighting factor $\alpha\approx2\omega_0$, where $\omega_0$ is the first eigen-frequency of the discretised system, approximated by the stiffness-mass Rayleigh ratio 
\begin{align} \label{SM3}
  {\omega_0}^2={\rm{max}\left(\dfrac{\bmsf{du}^{T}\left[\bmsf{T}_{k}-\bmsf{T}_{k-1}\right]}{\bmsf{du}^{T}\bmsf{M}_{\rm{fict}}\bmsf{du}},0\right)}\, ,
\end{align}
which has been modified specifically for the dynamic relaxation method \cite{oakley:1995}.

\section{Results}\label{sec3}


\subsection{Constitutive parameter calibration}\label{CPC}

\begin{figure}[!h]
    \centering
    \includegraphics[width=0.48\textwidth]{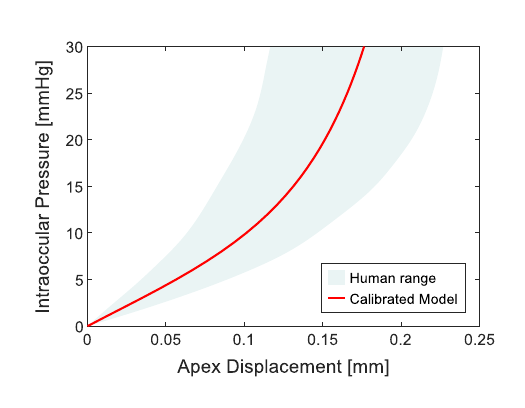}
    \caption{The fitted mechanical response of the presented multiscale model for the cornea under the action of the IOP, additionally displaying the range of biological variability in the ex vivo pressure inflation experimental data.}
    \label{fig:fig3}
\end{figure}

When computationally modelling soft biological tissues, patient-specific geometries can only confer physiologically predictive results by also establishing the corresponding patient-specific material parameters. Unfortunately, in the context of the human cornea, the lack of suitable in vivo mechanical tests that can differentiate the response of the cornea from that of the overall system (eyeball) renders the determination of the tissue's exact constitutive properties a challenging feat. However, to arrive at reasonable numerical results that are realistic to the human cornea, we start by calibrating the proposed multiscale model to experimental data concerning the ex vivo pressure inflation testing of the isolated cornea \cite{elsheikh:2010}. The testing protocol implies the complete blockage of the sclera adjacent to the limbus and the progressive increase of the applied IOP from 0 to 30~mmHg. A corneal geometry with discretisation parameters $N_L=3$ and $N_M=26$ was used for the fitting. The identified set of constitutive parameters is reported in Table~\ref{Tab2}, and the mechanical response reached is shown in Fig.~\ref{fig:fig3}, demonstrating that the model successfully captures the mean mechanical behaviour over the experimental range of variability. The calibrated parameters have been used in the numerical calculations described later.

\begin{table}[h]
\caption{Parameters obtained following the calibration of the model to human cornea pressure inflation testing data \cite{elsheikh:2010}, where discretisation parameters of $N_L=3$ and $N_M=26$ were used. The table also includes the calibrated parameters for the variance-based model used for comparative analysis \cite{pandolfi:2008}.}
\label{GeometricParameters}
\label{Tab2}
\begin{tabular}{l l l l}
\hline
\noalign{\vskip 0.15cm}
\hspace{0.2cm} & & \textbf{Value} & \textbf{Unit}\cr
\noalign{\vskip 0.1cm}
\hline
\noalign{\vskip 0.2cm}
\hspace{0.2cm}\scalebox{1.1}{\textbf{Coupled multiscale model}}\cr
\noalign{\vskip 0.2cm}
\hspace{0.6cm}\textbf{Truss elements}\cr
\noalign{\vskip 0.1cm}
\hspace{1.0cm}{Referential cross-sectional area}& $\overline{A}$ &1.00&$\rm{mm}^2$ \cr
\noalign{\vskip 0.1cm}
\hspace{1.0cm}\textit{Collagen fibril}\cr
\hspace{1.4cm}{Elastic Stiffness}& ${k}_{\rm{1}}$ &1.8&\rm{MPa}\cr
\hspace{1.4cm}{Rigidity Parameter}& ${k}_{\rm{2}}$ &4000&\cr
\noalign{\vskip 0.1cm}
\hspace{1.0cm}\textit{Proteoglycan crosslink}\cr
\hspace{1.4cm}{Minimum Potential energy}& ${\varepsilon}_{\rm{}}$ &0.01&\rm{MPa}\cr
\hspace{1.4cm}{Exponent}& $a$ & 6&  \cr
\noalign{\vskip 0.1cm}
\hspace{0.6cm}\textbf{Solid elements}\cr
\noalign{\vskip 0.1cm}
\hspace{1.0cm}\textit{ECM continuum material}\cr
\hspace{1.4cm}{Shear modulous 1}& ${\mu}_{1}$ &0.0015&\rm{MPa}\cr
\hspace{1.4cm}{Shear modulous 2}& ${\mu}_{2}$ &-0.0014&\rm{MPa}\cr
\hspace{1.4cm}{Bulk modulous}& ${K}$ &5&\rm{MPa}\cr
\noalign{\vskip 0.1cm}
\noalign{\vskip 0.2cm}
\hspace{0.2cm}\scalebox{1.1}{\textbf{Variance-based model (two fibril families)}}\cr
\noalign{\vskip 0.2cm}
\hspace{0.6cm}\textbf{Solid elements}\cr
\noalign{\vskip 0.1cm}
\hspace{1.0cm}\textit{Anisotropic collagen-related contribution}\cr
\hspace{1.4cm}{Fibril stiffness parameter (both families)}& ${k_1}$ &0.2&\rm{MPa}\cr
\hspace{1.4cm}{Fibril rigidity (both families)}& $k_2$ &510&\cr
\hspace{1.4cm}{Dispersion coefficient (both families)}& ${\kappa}$ &location dependant\cr
\noalign{\vskip 0.1cm}
\hspace{1.0cm}\textit{Isotropic contribution}\cr
\hspace{1.4cm}{Shear modulous 1}& ${\mu}_{1}$ &0.0015&\rm{MPa}\cr
\hspace{1.4cm}{Shear modulous 2}& ${\mu}_{2}$ &-0.0014&\rm{MPa}\cr
\hspace{1.4cm}{Bulk modulous}& ${K}$ &5&\rm{MPa}\cr
\noalign{\vskip 0.2cm}
\hline
\end{tabular}
\end{table}

\subsection{Influence of the geometric discretisation}

As the proposed model incorporates a combination of both truss and solid finite elements to upscale the microstructural features of corneal soft tissue to the macroscale, it is necessary to investigate how the geometric discretisation impacts the obtained numerical results.

\begin{figure}[!h]
    \centering
    \includegraphics[width=1.0\textwidth]{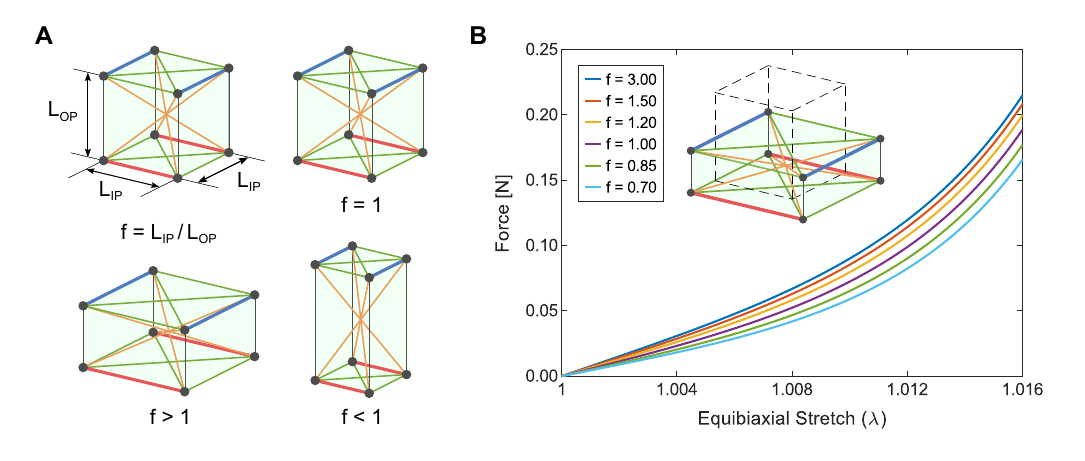}
    \caption{(A) Diagram illustrating the definition of the shape factor $f$, as well as representative examples of the referential state for the cases of $f=1$, $f>1$, and $f<1$. (B) The mechanical response of a single unit cell exposed to planar equibiaxial tension, for varying values of $f$, including a diagram showcasing the referential and deformed states for the case of $f=1$.}
    \label{fig:fig4}
\end{figure}

As depicted in Fig.~\ref{fig:fig4}A, a unit cell has three dimensions, an out-of-plane height $L_{OP}$, and two in-plane dimensions. The method of discretisation for the corneal geometry ensures that for all the generated unit cells, the two in-plane dimensions are always approximately equal, and so we use the average in-plane dimension $L_{IP}$ to represent both. A shape factor $f=L_{IP}/L_{OP}$ can then be defined, which provides pertinent information relating to the relative dimensions of a given unit cell, cf. \cite{koery:2024}. For instance, a value of $f=1$ refers to a referential cubic unit cell, whilst values of $f<1$ and $f>1$ refer to referential dimensions that are elongated in the out-of-plane and in-plane directions, respectively.

The influence of the shape factor $f$ upon the behaviour of a single unit cell exposed to planar equibiaxial tension, a loading circumstance closely resembling that of the unit cell in the full organ-level problem, can then be examined. The referential cross-sectional area of the facets over which the force is applied is kept the same for each test to ensure that the continuum ECM force contribution remains the same. From Fig.~\ref{fig:fig4}B, it can be seen that for decreasing values of $f$, the response softens, which is a consequence of the diagonal trusses within the unit cell representing the PG crosslinks. For different values of $f$, these trusses experience a different axial elongation for the same biaxial deformation of the unit cell, and as such, provide a different contribution to the overall force.

\begin{figure}[!h]
    \centering
    \includegraphics[width=1.0\textwidth]{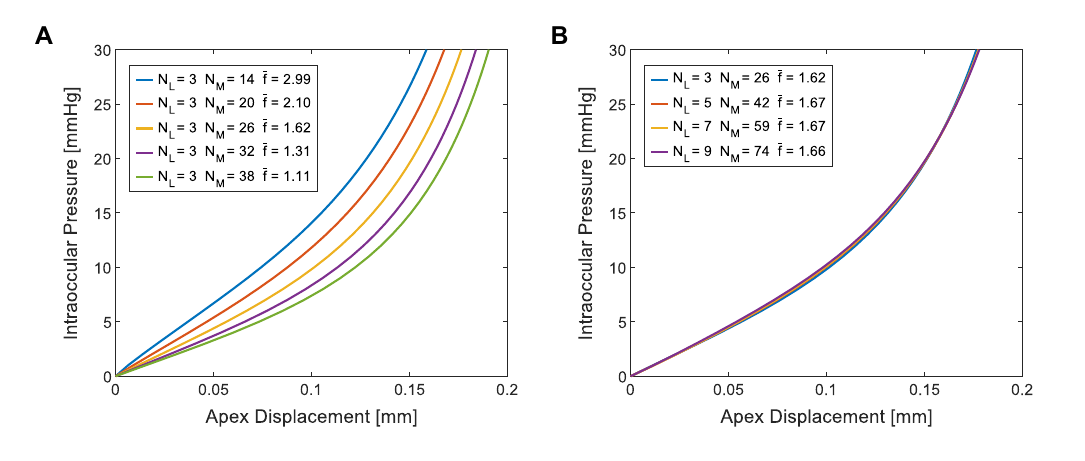}
    \caption{(A) Simulations for differing values of the discretisation parameters $N_L$ and $N_M$ that give varying magnitudes of $\bar{f}$ and therefore alternate mechanical behaviours. (B) Simulations for differing values of $N_L$ and $N_M$ that are specifically chosen to give approximately equal magnitudes of $\bar{f}$ and therefore correspondingly equivalent mechanical behaviour.}
    \label{fig:fig5}
\end{figure}

Next, we focus our attention on the effect of the discretisation parameters, $N_L$ and $N_M$, on the mechanical behaviour of the pressurised cornea, as the choice of said parameters will obviously dictate the shape of the unit cells that make up the corneal geometry as a whole. This impact is measured through the average value of the shape factor $\bar{f}$, across all resident unit-cells, allowing for an effective comparison of the different degrees of discretisation. From Fig.~\ref{fig:fig5}A it can be seen that the influence of the relative dimensions observed for a single unit cell, also manifests at the whole organ-level. The number of layers is held constant at $N_L=3$, whilst the value of $N_M$ is altered. For more elements across the meridian diameter, the magnitude of $\bar{f}$ decreases, and there is a corresponding reduction in the cornea's macroscopic stiffness. From Fig.~\ref{fig:fig5}B it is evident that if $N_L$ and $N_M$ are chosen to provide values of $\bar{f}$ that are approximately equal in magnitude, the mechanical behaviour of the cornea is relatively consistent, with the minor discrepancies also likely attributable to standard mesh convergence phenomena.

\subsection{Application of the model to Keratoconus}

The proposed coupled multiscale model has the potential to describe the mechanical outcome of pathologies such as keratoconus, which are characterised by a reduction in material stiffness within a localised area of the cornea, leading to the development of a distinct conical shape. To assess the model's ability to simulate keratoconus, we compare its prediction with that of a phenomenological variance-based continuum model of the corneal stroma, previously applied to the simulation of mechanical tests and refractive surgery procedures \cite{pandolfi:2012}. It accounts for the reinforcement of the collagenous matrix via two families of dispersed collagen fibrils, obeying a von Mises orientation distribution governed by a scalar parameter $b(\bx)$, itself a function of the spatial location within the cornea. The model employs the average and variance of the fourth pseudo-invariant $\Iqs$, coupling the orientation of a single fibril to the deformation gradient. As with the coupled multiscale model, we calibrate its material parameters to experimental pressure inflation data (Table~\ref{Tab2}), such that the mechanical response of both models are approximately equivalent for the healthy eye. Note that both models use the Mooney-Rivlin strain energy function to describe the isotropic ECM material. A brief overview of the variance-based model is provided in Appendix~\ref{AppB}.

To model the degeneration of the corneal stroma, the various stiffness parameters of both models are reduced according to a scalar damage field $d \in \{0,1\}$. Clearly, the reduction is not applied to the bulk modulous $K$, which acts as a penalty coefficient to enforce incompressibility. The spatial distribution of $d$ is assumed to be quadratic and to cover a circular area, with maximum tissue degeneration ($d=1$) at the centre and no degeneration ($d=0$) at the boundary of the damaged zone, cf.~\cite{debellis:2023}. The centre of the allocated damage field is situated 1~mm towards the inferior side of the cornea, along the SI meridian, in agreement with clinical evidence indicating that keratoconus tends to localise in the lower portion of the corneal surface~\cite{rabinowitz:1998}. 

\begin{figure}[!h]
    \centering
    \includegraphics[width=0.48\textwidth]{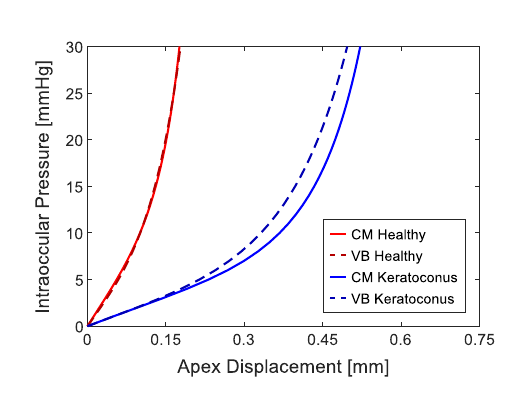}
    \caption{The mechanical response of the presented Coupled multiscale (CM) model and the Variance-based (VB) Model for both the healthy (calibrated to experimental data \cite{elsheikh:2010}) and keratoconus case.}
    \label{fig:fig6}
\end{figure}

The global mechanical response under the action of the IOP, for the proposed coupled multiscale model and the variance-based model, applied to the health and keratoconus cases, is shown in Fig.~\ref{fig:fig6}. Whilst the two models predict the same behaviour under healthy conditions, they differ for the keratoconus case, with the coupled multiscale model exhibiting greater compliance.

\begin{figure}[!h]
    \centering
\includegraphics[width=1.0\textwidth]{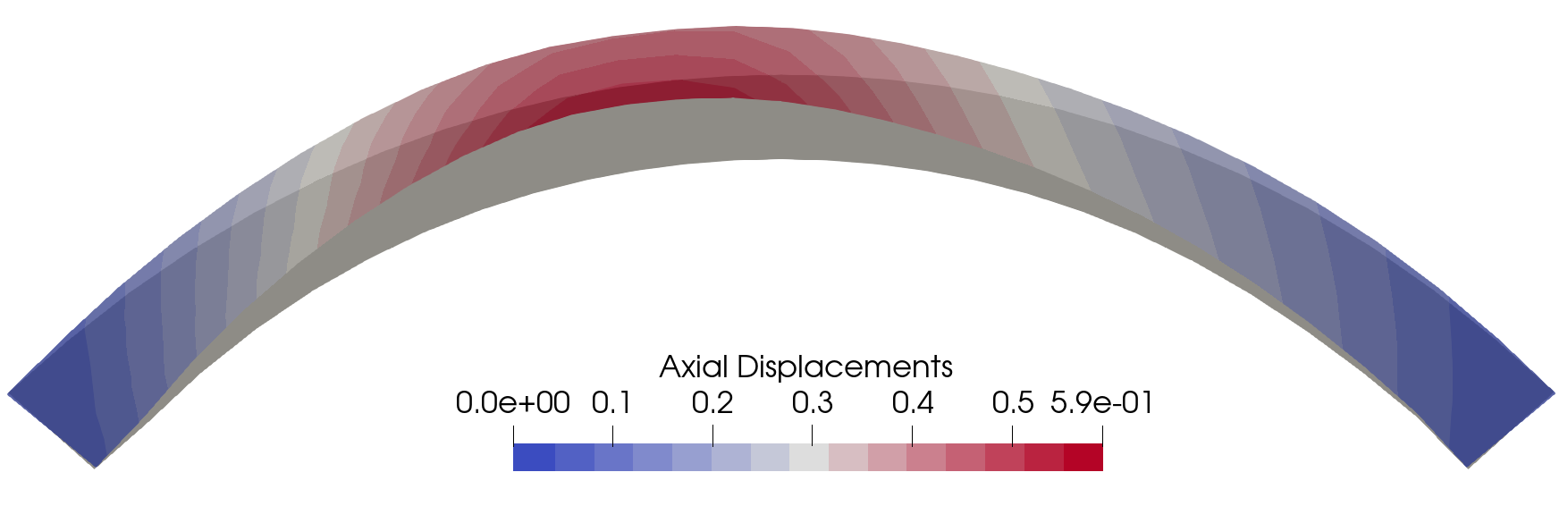}
    \caption{SI meridional profile of the cornea under healthy and keratoconus conditions obtained with the coupled multiscale model. The keratoconus configuration is highlighted with a colour map visualising the displacements along the optic axis. A conical protrusion has formed, characteristic of keratoconus, with a localised narrowing at the location of maximum damage.}
    \label{fig:fig7}
\end{figure}

\begin{figure}[!h]
    \centering
\includegraphics[width=1.0\textwidth]{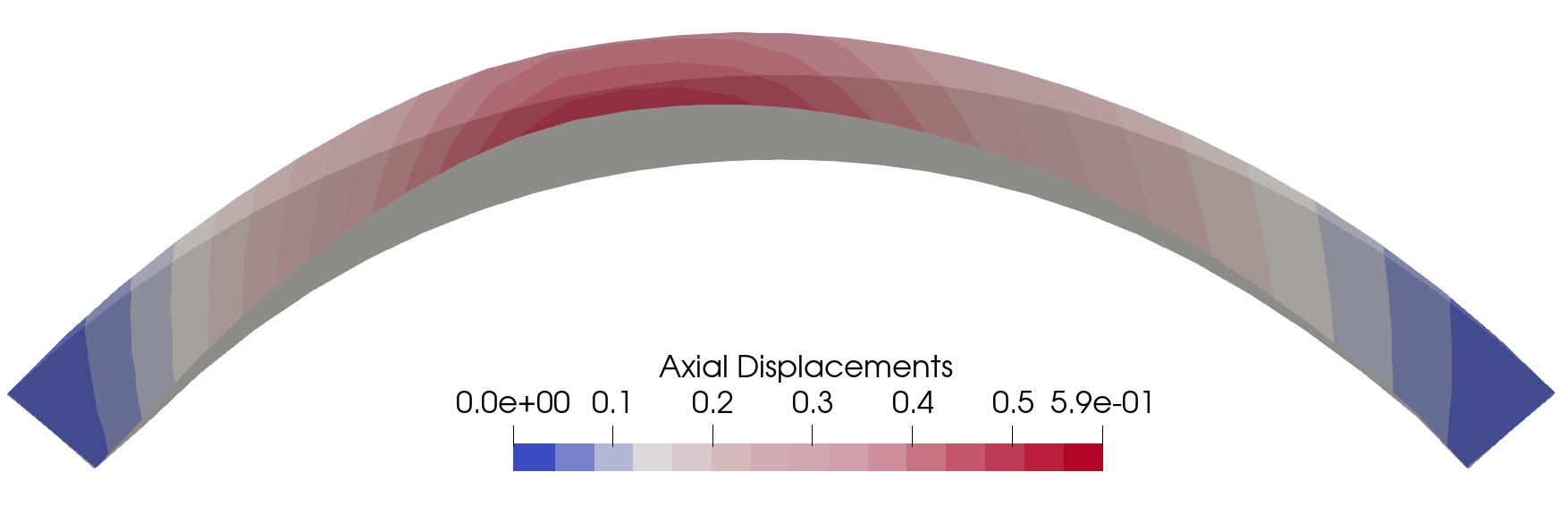}
    \caption{SI meridional profile of the cornea under healthy and keratoconus conditions obtained with the variance-based continuum model. The keratoconus configuration is highlighted with a colour map visualising the displacements along the optic axis. Whilst a localised bulging occurs, the deformation is less pronounced than that obtained with the coupled multiscale model.}
    \label{fig:fig8}
\end{figure}

SI meridian sections of the cornea are displayed in Figs.~\ref{fig:fig7}-\ref{fig:fig8}, demonstrating that both models are able to capture the key features of the pathological configuration. Specifically, Fig.~\ref{fig:fig6} compares the profiles of the healthy and keratoconus corneas obtained with the coupled multiscale model, whilst Fig.~\ref{fig:fig7} compares the same profiles obtained with the variance-based model. 

\begin{figure}[!h]
    \centering
\includegraphics[width=1.0\textwidth]{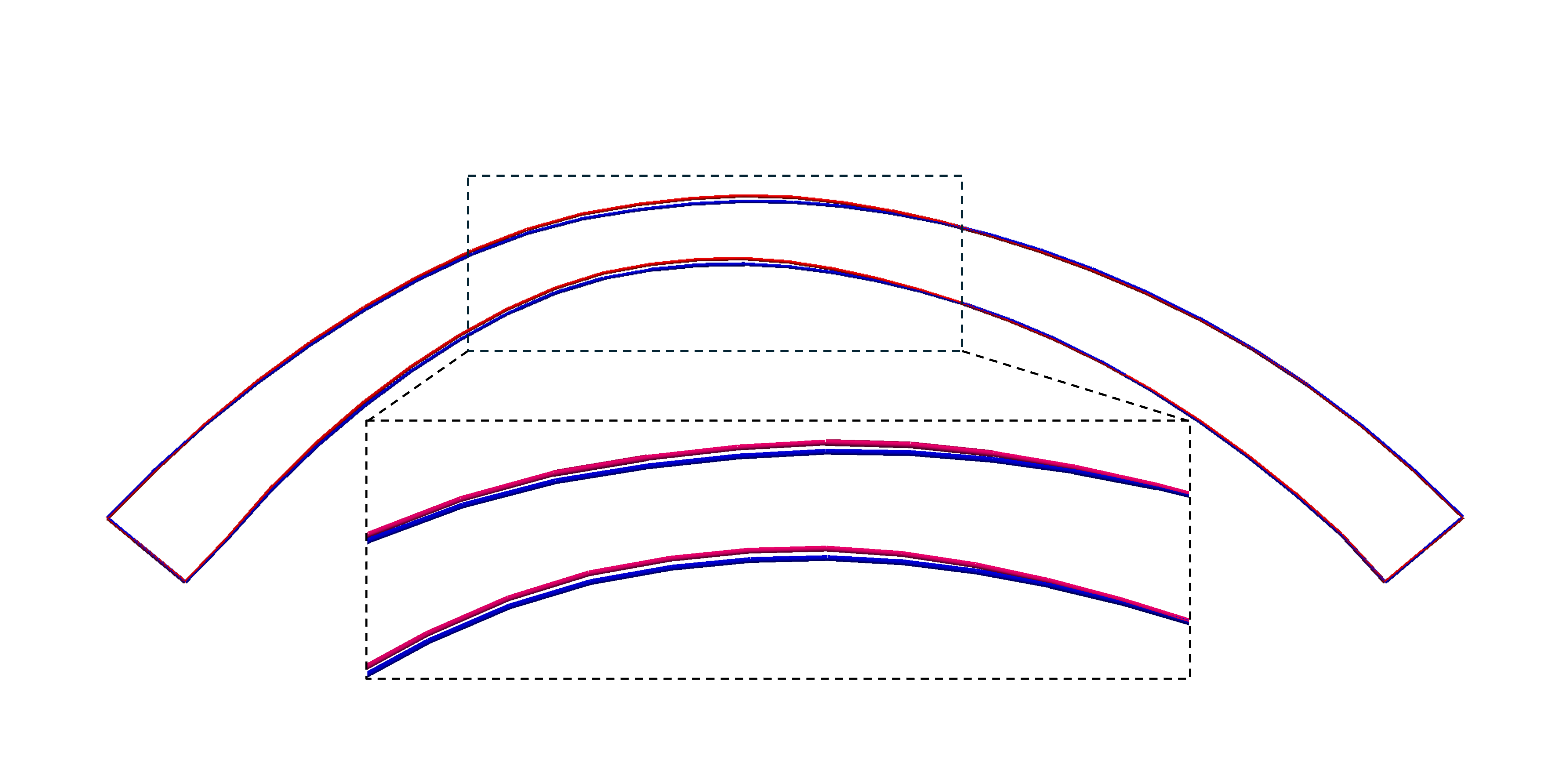}
    \caption{Direct comparison of the SI profile of the keratoconus corneas at physiological IOP (15 mmHg), obtained with the coupled multiscale and the variance-based models.}
    \label{fig:fig9}
\end{figure}

A direct comparison between the keratoconus profiles predicted by the two models is provided in Fig.~\ref{fig:fig9}. The conical protrusion appears slightly more pronounced in the coupled multiscale model, which reveals $\approx$ 7\% higher displacements at the corneal apex. Moreover, a more pronounced thinning is observed in the region of the conus apex, suggesting a more marked stiffness reduction.

\section{Discussion and conclusions}\label{sec4}

In recent times, the computational modelling of biological systems has become a vital tool supporting the design of surgical and pharmacological treatments. Despite the difficulties associated with the selection of in vivo physical properties, numerical models can, for instance, be used to simulate surgical procedures and augment clinical awareness regarding optimal outcomes, especially when the surgery is risky. They are also becoming increasingly relevant in the study of disease, shedding light on the myriad of factors linked to the incidence and progression of corneal pathologies. In silico modelling could therefore be the key to solving, or at least partially alleviating, the various mysteries of corneal biomechanics.

Advanced continuum-based numerical models of the human cornea that account for tissue anisotropy have been successfully deployed in the simulation of healthy physiological states \cite{pandolfi:2006}, in vivo mechanical tests \cite{arizagracia:2015, simonini:2016b, montanino:2023}, and surgical procedures \cite{sanchez:2014, montanino:2023}. Nevertheless, such approaches, which disregard the underlying microstructure, have proven to be insufficient in the modelling of tissue degeneration. \cite{pandolfi:2006, simonini:2022}. Recently, a micromechanical approach upscaled to the organ level has been successfully applied to model the evolution of keratoconus \cite{pandolfi:2019, pandolfi:2023}. The approach explicitly accounts for the mechanical interaction of salient features such as fibrous collagen and associated crosslinks by characterising them as structural truss elements within a spatially repeating unit-cell, which defines the load-carrying lamellar structure of the stroma, thus providing a more physical interpretation of the microstructure compared to entirely phenomenological continuum models. 

The study presented here has been devoted to the extension of this existing discrete framework to further include a continuum representation of the non-collagenous ECM in which the mechanically significant constituents are embedded. With the framework now amended to treat the cornea as a solid entity, this multiscale approach now correctly enforces material incompressibility, a hallmark feature of soft biological tissues. It is also well-placed for the possible integration of mathematical descriptions accounting for coupled phenomena such as heat conduction during surgical treatments and fluid/ion fluxes occurring at the posterior surface to preserve corneal transparency. A new finite element truss formulation has also been designed and incorporated, which properly reflects the large deformation and non-linearity of structural proteins represented as truss elements. Importantly, the superposition of the collagen trusswork and the ECM continuum within a finite element discretisation is computationally straightforward, and the amalgamation of the two approaches has the potential to overcome the fundamental failings of either alone.

Clearly, a new modelling approach necessitated a revisitation of the material parameters, which have been calibrated to experimental ex vivo data taken from the literature, as well as informed by our past modelling endeavours. An in-depth analysis concerning the effect of the cornea's geometric discretisation was also carried out, revealing an inherent mesh-dependency of the model arising from the intra-laminar and inter-laminar crosslinks that form a unit cell. However, most significantly, we have demonstrated that the proposed model, whilst using relatively simplistic numerical methods, albeit in a novel and innovative way, is sufficiently advanced for describing corneal pathologies such as ectasia and keratoconus. By applying a spatial reduction in the stiffness of all tissue constituents, it was possible to simulate localised mechanical instability leading to the formation of a conus, to an extent that more closely resembles clinical observations in comparison to previously reported, purely continuum-based approaches.

Surprisingly, in the present study, we have additionally been able to obtain favourable results for the variance-based model when applied to the simulation of keratoconus in the cornea, predicting a satisfactory shape for the conus. This can be explained by the fact that, with the support of the couple multiscale model, we have found constitutive parameters somewhat different from those used in previous studies, cf.~\cite{pandolfi:2024}. Specifically, the proportions of the various mechanical contributions shifted towards a heightened stiffness of the collagen fibrils relative to the isotropic ECM material, following the indications obtained from the coupled multiscale model. However, both the proposed multiscale model and the variance-based model cannot suitably capture the thinning of the cornea (50\% based on clinical observations) to the extent that the trusswork alone is able to characterise \cite{pandolfi:2023}. This issue is clearly related to the incompressibility constraint of the hyperelastic solid elements, which is not necessarily applicable to the progressive degeneration of the tissue in its diseased state.

There are, additionally, certain limitations associated with the current model and its implementation. For one, whilst the number of degrees of freedom is unchanged compared to conventional continuum approaches, the large number of truss and solid elements for even moderately discretised geometries can lead to significant computational expense and lengthy simulation times. However, this issue is partially mitigated through concurrent computing, as by using the dynamic relaxation solver, each element is managed independently of the others. Another shortcoming is that the model does not consider the biological and chemical aspects associated with keratoconus and other diseases, such as endothelial dysregulation and corneal swelling. This may be remedied by replacing the hyperelastic description of the ECM material with a biphasic description capable of modelling such phenomena and correctly accounting for thinning (volumetric reduction), which is currently under development. Lastly, as has already been alluded to, the mathematical portrayal of the microstructure is generalised, representing only a fraction of the lamellae present in the real cornea. Computational limitations at the present time make it infeasible to model the entire collagenous architecture of the cornea in the manner presented here. The behaviour of a multilayered trusswork was recently investigated for an increasing number of laminar layers, leading to the definition of a continuum equivalent anisotropic material \cite{koery:2024}. Future strategies may also rely on the advent of sophisticated physics-based machine learning techniques to model the lamellar structure and effectively bridge different length-scales.

Finally, the need for patient-specific material properties represents an ongoing challenge if the effective numerical modelling of the cornea for various clinical and research-based applications is to be fully realised, a factor that depends greatly upon the availability of data garnered from in vivo testing methods, which at the present time remains a bottleneck in the modelling of not just the cornea, but all load-bearing soft biological tissues. 

\section*{Acknowledgements}
This work was supported by the Italian Ministry of University and Research (MUR) under the PRIN 2022 project CORTIS (Grant No. 2022TWKA72). AP and MLDB are grateful for the support of the Italian National Group of Physics-Mathematics (GNFM) belonging to the Italian National Institution of High Mathematics ``Francesco Severi'' (INDAM).

\backmatter

\begin{appendices}

\section{Truss Element Tangent Stiffness Matrix Derivation}\label{AppA}

The derivation of the tangent stiffness matrix relating to a nonlinear large deformation truss element, first introduced during the discussion of the formulation in Section~(\ref{sec2.4.1}), is an alternative form of that presented in \cite{bonet:2008}. To begin with, the directional derivative of the current length vector, $l\bn=(\bx_b-\bx_a)$, is determined as
\begin{align}\nonumber
    D\left[\bx_b-\bx_a\right][\bu]
    =\
    \dfrac{d}{d\epsilon}\bigg|_{\epsilon=0}
    (\bx_b+\epsilon\bu_b-\bx_a-\epsilon\bu_a)=(\bu_b-\bu_a)\, ,
\end{align}
which is based on the definitions of the scalar length of the truss, $l=\left\{(\bx_b-\bx_a)\cdot(\bx_b-\bx_a)\right\}^{1/2}$, and the unit vector, $\bn=(\bx_b-\bx_a)/l$, describing its direction in space. We can then determine the directional derivatives of several quantities required for the subsequent derivation, such as,
\begin{align}\nonumber
\begin{array}{rl}
    D\left[l^{-1}(\bx)\right][\bu] = -l^{-2}\bn\cdot(\bu_b-\bu_a)\, , \\[10pt] 
    D\left[\lambda(\bx)\right][\bu]=\dfrac{1}{L}\bn\cdot(\bu_b-\bu_a)\, ,
\end{array}
\end{align}
where, $\lambda=l/L$, is the non-dimensional stretch of the deformed truss. The directional derivative of the truss's internal force at Node $\rm{b}$ with respect to the elements displacement, ${\bu=[\bu_{\rm{a}}, \bu_{\rm{b}}]}^{T}$, following on from the nodal internal force definitions in Eq.~(\ref{Truss1}), and the application of the aforementioned expressions, is then derived as
\begin{align}\nonumber
\begin{array}{rl}
  &D\left[\bT^{}_b(\lambda(\bx),\bn(\bx))\right] [\bu] \\[20pt] =& 
  D\left[P(\lambda(\bx))\right][\bu]A\bn + P(\lambda)D\left[A\bn(\bx)\right][\bu]\\[20pt]
  =
  &\dfrac{dP(\lambda)}{d\lambda}D\left[\lambda(\bx)\right][\bu]A\bn(\bx) +P(\lambda)A\left(D\left[l^{-1}(\bx)\right][\bu](\bx_b-\bx_a)+\dfrac{1}{l}D\left[\bx_b-\bx_a\right][\bu]\right)\\[20pt]
  =
  &\dfrac{\mathcal{A}(\lambda)A}{L}\bn\cdot(\bu_b-\bu_a)\bn
  +P(\lambda)AlD\left[l^{-1}(\bx)\right][\bu]\bn+\dfrac{P(\lambda)A}{l}\left(\bu_b-\bu_a\right)\\[20pt]
  =
  &\dfrac{\mathcal{A}(\lambda)A}{L}\bn\cdot(\bu_b-\bu_a)\bn-\dfrac{P(\lambda)A}{l}\bn\cdot(\bu_b-\bu_a)\bn+\dfrac{P(\lambda)A}{l}\left(\bu_b-\bu_a\right)\\[20pt]
  =
  &\left(\dfrac{\mathcal{A}(\lambda)A}{L}-\dfrac{P(\lambda)A}{l}\right) (\bn\otimes\bn)\left(\bu_b-\bu_a\right)+\dfrac{P(\lambda)A}{l}(\bu_b-\bu_a)\\[20pt]
  =
  &[\alpha(\lambda)-\beta(\lambda)](\bn \otimes\bn)\left(\bu_b-\bu_a\right)+\beta(\lambda)\bI\left(\bu_b-\bu_a\right)\, ,
\end{array}
\end{align}
which, given that, $D\left[\bT_a(\bx)\right][\bu]=-D\left[\bT_b(\bx)\right][\bu]$, due to Eq.~(\ref{Truss1}), gives rise to the matrix representation of the element tangent stiffness matrix detailed in Eq.~(\ref{Truss2}) with the individual contributions provided in Eq.~(\ref{Truss3}).

\section{Variance-based continuum model of corneal tissue}\label{AppB}

A brief overview of the key aspects relating to the variance-based model is provided here; however, for a more expansive detailing, the reader is referred to \cite{pandolfi:2012}. The model consists of two families of spatially dispersed collagen fibrils, such that the cornea is represented as a reinforced anisotropic material, where the specific orientation of fibrils follows the most advanced findings in the literature \cite{aghamohammadzadeh:2004}. The total strain energy consists of three terms
\begin{align}
    \Psi = \PV + \PI + \PA \,,
    \nonumber
\end{align}
with $\PV$ denoting the volumetric strain energy which takes the form
\begin{align}
    \PV(J)
    =
    \frac{1}{4}\, K \,(J^2 - 1 -2 \log{J}), 
    \nonumber
\end{align}
where $J>0$ is the determinant of the deformation gradient $\bF$, and $K$ a penalization coefficient analogous to the bulk modulous. The isotropic energy $\PI$ follows the Mooney-Rivlin model and is defined as
\begin{equation}
\label{eq:PsiI}
    \PI(\overline I_1, \overline I_2)
    =
    \frac{1}{2} \mu_1 (\overline{I}_1 -3)
    +
    \frac{1}{2} \mu_2 (\overline{I}_2 -3) \, ,
    \quad
    \Iu = \tr \, {\overline \bC} \, , 
    \quad 
    {\overline I_2} = \frac{1}{2} [ ({\tr \, {\overline \bC}})^2 - \tr (\overline \bC^2) ] \, ,
    \nonumber
\end{equation}
where $\mu = \mu_1 + \mu_2$ is the elastic shear modulous of the material. The terms $\Iu$ and $\overline I_2$ denote the first and the second invariants, respectively, of the isochoric Cauchy-Green deformation tensor $\overline{\bC} = \overline{\bF}^T\overline{\bF}$, with $\overline{\bF} = J^{-1/3}\bF$. The anisotropic strain energy $\PA$ describing the contribution of the two collagen fibril families is
\begin{equation}
\label{eq:PsiAbar}
    \PA (\Iqs_M, \sig_M)
    =
    \sum_{M=1}^2 \frac{k_{1M}}{2k_{2M}}
    \exp \left[k_{2M}\left(\Iqs_M - 1 \right)^2\right] \left(1 + K_M^*(\Iqs_M) \sig_M \right),
    \nonumber
\end{equation}
where $k_{1M}$ is a stiffness parameter, controlling the fibril behaviour at moderate deformations, and $k_{2M}$ is a dimensionless rigidity parameter, regulating the response at large deformations. The pseudo-invariants $\Iqs_M$ are defined as
\begin{equation}\label{eq:Iqs}
    \Iqs_M
    =
    \bH_M : \Co \, , \quad
    \bH_M
    =
    \langle \bA_M \otimes \bA_M \rangle =
    \kappa_M \bI + (1 - 3 \kappa_M) \bA_{M0} \, , 
    \quad \bA_M = \ba_M \otimes \ba_M \, ,
    \nonumber
    \nonumber
\end{equation}
where $\langle \bullet \rangle$ denotes the average over the unit sphere, $\bA_{M0}=\ba_0 \otimes \ba_0$ refers to the main orientation of the fibril distribution, and the scalar parameter $\kappa_M$ is defined according to
\begin{equation}\label{eq:kappaKappaHat}
    \kappa_M
    =
    \frac{1}{4}\int_0^\pi\rho_M(\Theta)\sin^3\Theta d\Theta \, ,
    \nonumber
\end{equation}
where $\rho_M(\Theta)$ is the spatial probability distribution (e.g., von Mises). Finally, the relations
\begin{equation}\label{eq:Kappa}
    \sig_M
    =
    \Co : \langle \bA_M \otimes \bA_M \rangle : \Co - \big(\bH_M :\Co \big)^2 \, ,
    \quad
    K_M^*(\Iqs_M)
    =
    k_{2M} +  2 \,k_{2M}^2 \, \left(\Iqs_M - 1\right)^2 
    \nonumber
\end{equation} 
provide the variance contribution to $\PA$ and its amplification coefficient, respectively. 
\end{appendices}

\bibliography{microcornea}


\begin{thebibliography}{30}
\ifx \bisbn   \undefined \def \bisbn  #1{ISBN #1}\fi
\ifx \binits  \undefined \def \binits#1{#1}\fi
\ifx \bauthor  \undefined \def \bauthor#1{#1}\fi
\ifx \batitle  \undefined \def \batitle#1{#1}\fi
\ifx \bjtitle  \undefined \def \bjtitle#1{#1}\fi
\ifx \bvolume  \undefined \def \bvolume#1{\textbf{#1}}\fi
\ifx \byear  \undefined \def \byear#1{#1}\fi
\ifx \bissue  \undefined \def \bissue#1{#1}\fi
\ifx \bfpage  \undefined \def \bfpage#1{#1}\fi
\ifx \blpage  \undefined \def \blpage #1{#1}\fi
\ifx \burl  \undefined \def \burl#1{\textsf{#1}}\fi
\ifx \doiurl  \undefined \def \doiurl#1{\url{https://doi.org/#1}}\fi
\ifx \betal  \undefined \def \betal{\textit{et al.}}\fi
\ifx \binstitute  \undefined \def \binstitute#1{#1}\fi
\ifx \binstitutionaled  \undefined \def \binstitutionaled#1{#1}\fi
\ifx \bctitle  \undefined \def \bctitle#1{#1}\fi
\ifx \beditor  \undefined \def \beditor#1{#1}\fi
\ifx \bpublisher  \undefined \def \bpublisher#1{#1}\fi
\ifx \bbtitle  \undefined \def \bbtitle#1{#1}\fi
\ifx \bedition  \undefined \def \bedition#1{#1}\fi
\ifx \bseriesno  \undefined \def \bseriesno#1{#1}\fi
\ifx \blocation  \undefined \def \blocation#1{#1}\fi
\ifx \bsertitle  \undefined \def \bsertitle#1{#1}\fi
\ifx \bsnm \undefined \def \bsnm#1{#1}\fi
\ifx \bsuffix \undefined \def \bsuffix#1{#1}\fi
\ifx \bparticle \undefined \def \bparticle#1{#1}\fi
\ifx \barticle \undefined \def \barticle#1{#1}\fi
\bibcommenthead
\ifx \bconfdate \undefined \def \bconfdate #1{#1}\fi
\ifx \botherref \undefined \def \botherref #1{#1}\fi
\ifx \url \undefined \def \url#1{\textsf{#1}}\fi
\ifx \bchapter \undefined \def \bchapter#1{#1}\fi
\ifx \bbook \undefined \def \bbook#1{#1}\fi
\ifx \bcomment \undefined \def \bcomment#1{#1}\fi
\ifx \oauthor \undefined \def \oauthor#1{#1}\fi
\ifx \citeauthoryear \undefined \def \citeauthoryear#1{#1}\fi
\ifx \endbibitem  \undefined \def \endbibitem {}\fi
\ifx \bconflocation  \undefined \def \bconflocation#1{#1}\fi
\ifx \arxivurl  \undefined \def \arxivurl#1{\textsf{#1}}\fi
\csname PreBibitemsHook\endcsname

\bibitem[\protect\citeauthoryear{Mishima and Hedbys}{1968}]{mishima:1968}
\begin{barticle}
\bauthor{\bsnm{Mishima}, \binits{S.}},
\bauthor{\bsnm{Hedbys}, \binits{B.O.}}:
\batitle{Physiology of the cornea}.
\bjtitle{International ophthalmology clinics}
\bvolume{8}(\bissue{3}),
\bfpage{527}--\blpage{560}
(\byear{1968})
\end{barticle}
\endbibitem

\bibitem[\protect\citeauthoryear{Bron}{2001}]{bron:2001}
\begin{barticle}
\bauthor{\bsnm{Bron}, \binits{A.J.}}:
\batitle{The architecture of the corneal stroma}.
\bjtitle{{British Journal of Ophthalmology}}
\bvolume{85},
\bfpage{379}--\blpage{383}
(\byear{2001})
\end{barticle}
\endbibitem

\bibitem[\protect\citeauthoryear{Maurice}{1957}]{maurice:1957}
\begin{barticle}
\bauthor{\bsnm{Maurice}, \binits{D.M.}}:
\batitle{The structure and transparency of the cornea}.
\bjtitle{The Journal of physiology}
\bvolume{136}(\bissue{2}),
\bfpage{263}
(\byear{1957})
\end{barticle}
\endbibitem

\bibitem[\protect\citeauthoryear{Abahussin et~al.}{2009}]{abahussin:2009}
\begin{barticle}
\bauthor{\bsnm{Abahussin}, \binits{M.}},
\bauthor{\bsnm{Hayes}, \binits{S.}},
\bauthor{\bsnm{Cartwright}, \binits{N.E.K.}},
\bauthor{\bsnm{Kamma-Lorger}, \binits{C.S.}},
\bauthor{\bsnm{Khan}, \binits{Y.}},
\bauthor{\bsnm{Marshall}, \binits{J.}},
\bauthor{\bsnm{Meek}, \binits{K.M.}}:
\batitle{3{D} collagen orientation study of the human cornea using {X}-ray
  diffraction and femtosecond laser technology}.
\bjtitle{{Investigative Ophthalmology \& Visual Science}}
\bvolume{50}(\bissue{11}),
\bfpage{5159}--\blpage{5164}
(\byear{2009})
\doiurl{10.1167/iovs.09-3669}
\end{barticle}
\endbibitem

\bibitem[\protect\citeauthoryear{Meek et~al.}{1987}]{meek:1987}
\begin{barticle}
\bauthor{\bsnm{Meek}, \binits{K.M.}},
\bauthor{\bsnm{Blamires}, \binits{T.}},
\bauthor{\bsnm{Elliot}, \binits{G.F.}},
\bauthor{\bsnm{Gyi}, \binits{T.J.}},
\bauthor{\bsnm{Nave}, \binits{C.}}:
\batitle{The organization of collagen fibrils in the human corneal stroma: a
  synchroton x-ray diffraction study}.
\bjtitle{{Current Eye Research}}
\bvolume{6},
\bfpage{841}--\blpage{846}
(\byear{1987})
\end{barticle}
\endbibitem

\bibitem[\protect\citeauthoryear{Aghamohammadzadeh
  et~al.}{2004}]{aghamohammadzadeh:2004}
\begin{barticle}
\bauthor{\bsnm{Aghamohammadzadeh}, \binits{H.}},
\bauthor{\bsnm{Newton}, \binits{R.H.}},
\bauthor{\bsnm{Meek}, \binits{K.M.}}:
\batitle{X-ray scattering used to map the preferred collagen orientation in the
  human cornea and limbus}.
\bjtitle{{Structure}}
\bvolume{12}(\bissue{2}),
\bfpage{249}--\blpage{256}
(\byear{2004})
\end{barticle}
\endbibitem

\bibitem[\protect\citeauthoryear{Rabinowitz}{1998}]{rabinowitz:1998}
\begin{barticle}
\bauthor{\bsnm{Rabinowitz}, \binits{Y.S.}}:
\batitle{Keratoconus}.
\bjtitle{{Survey of Ophthalmogy}}
\bvolume{42},
\bfpage{297}--\blpage{319}
(\byear{1998})
\end{barticle}
\endbibitem

\bibitem[\protect\citeauthoryear{Simonini et~al.}{2022}]{simonini:2022}
\begin{barticle}
\bauthor{\bsnm{Simonini}, \binits{I.}},
\bauthor{\bsnm{Ni~Annaidh}, \binits{A.}},
\bauthor{\bsnm{Pandolfi}, \binits{A.}}:
\batitle{Numerical estimation of stress and refractive power maps in healthy
  and keratoconus eyes}.
\bjtitle{Journal of the Mechanical Behavior of Biomedical Materials}
\bvolume{131},
\bfpage{105252}
(\byear{2022})
\end{barticle}
\endbibitem

\bibitem[\protect\citeauthoryear{Simonini and Pandolfi}{2016}]{simonini:2016b}
\begin{barticle}
\bauthor{\bsnm{Simonini}, \binits{I.}},
\bauthor{\bsnm{Pandolfi}, \binits{A.}}:
\batitle{The influence of intraocular pressure and air jet pressure on corneal
  contactless tonometry tests}.
\bjtitle{{Journal of the Mechanical Behavior of Medical Biomaterials}}
\bvolume{58},
\bfpage{75}--\blpage{89}
(\byear{2016})
\doiurl{10.1016/j.jmbbm.2015.07.030}
\end{barticle}
\endbibitem

\bibitem[\protect\citeauthoryear{S\'anchez et~al.}{2014}]{sanchez:2014}
\begin{barticle}
\bauthor{\bsnm{S\'anchez}, \binits{P.}},
\bauthor{\bsnm{Moutsouris}, \binits{K.}},
\bauthor{\bsnm{Pandolfi}, \binits{A.}}:
\batitle{Biomechanical and optical behavior of human corneas before and after
  photorefractive keratectomy}.
\bjtitle{{Journal of Cataract \& Refractive Surgery}}
\bvolume{40}(\bissue{6}),
\bfpage{905}--\blpage{917}
(\byear{2014})
\end{barticle}
\endbibitem

\bibitem[\protect\citeauthoryear{Pandolfi and
  Manganiello}{2006}]{pandolfi:2006}
\begin{barticle}
\bauthor{\bsnm{Pandolfi}, \binits{A.}},
\bauthor{\bsnm{Manganiello}, \binits{F.}}:
\batitle{A material model for the human cornea. constitutive behavior and
  numerical analysis}.
\bjtitle{{Biomechanics and Modelling in Mechanobiology}}
\bvolume{5},
\bfpage{237}--\blpage{246}
(\byear{2006})
\end{barticle}
\endbibitem

\bibitem[\protect\citeauthoryear{Pandolfi and Vasta}{2012}]{pandolfi:2012}
\begin{barticle}
\bauthor{\bsnm{Pandolfi}, \binits{A.}},
\bauthor{\bsnm{Vasta}, \binits{M.}}:
\batitle{Fiber distributed hyperelastic modeling of biological tissues}.
\bjtitle{{Mechanics of Materials}}
\bvolume{44},
\bfpage{151}--\blpage{162}
(\byear{2012})
\end{barticle}
\endbibitem

\bibitem[\protect\citeauthoryear{Miller and Gasser}{2021}]{miller:2021}
\begin{barticle}
\bauthor{\bsnm{Miller}, \binits{C.}},
\bauthor{\bsnm{Gasser}, \binits{T.C.}}:
\batitle{A microstructurally motivated constitutive description of collagenous
  soft biological tissue towards the description of their non-linear and
  time-dependent properties}.
\bjtitle{Journal of the Mechanics and Physics of Solids}
\bvolume{154},
\bfpage{104500}
(\byear{2021})
\doiurl{10.1016/j.jmps.2021.104500}
\end{barticle}
\endbibitem

\bibitem[\protect\citeauthoryear{Miller and Gasser}{2022}]{miller:2022}
\begin{barticle}
\bauthor{\bsnm{Miller}, \binits{C.}},
\bauthor{\bsnm{Gasser}, \binits{T.C.}}:
\batitle{A bottom-up approach to model collagen fiber damage and failure in
  soft biological tissues}.
\bjtitle{Journal of the Mechanics and Physics of Solids}
\bvolume{169},
\bfpage{105086}
(\byear{2022})
\doiurl{10.1016/j.jmps.2022.105086}
\end{barticle}
\endbibitem

\bibitem[\protect\citeauthoryear{Pandolfi et~al.}{2019}]{pandolfi:2019}
\begin{barticle}
\bauthor{\bsnm{Pandolfi}, \binits{A.}},
\bauthor{\bsnm{Gizzi}, \binits{A.}},
\bauthor{\bsnm{Vasta}, \binits{M.}}:
\batitle{A microstructural model of cross-link interaction between collagen
  fibrils in the human cornea}.
\bjtitle{Philosophical Transactions of the Royal Society A}
\bvolume{377}(\bissue{2144}),
\bfpage{20180079}
(\byear{2019})
\end{barticle}
\endbibitem

\bibitem[\protect\citeauthoryear{Pandolfi et~al.}{2023}]{pandolfi:2023}
\begin{barticle}
\bauthor{\bsnm{Pandolfi}, \binits{A.}},
\bauthor{\bsnm{De~Bellis}, \binits{M.L.}},
\bauthor{\bsnm{Gizzi}, \binits{A.}},
\bauthor{\bsnm{Vasta}, \binits{M.}}:
\batitle{Modeling the degeneration of the collagen architecture in a
  microstructural model of the human cornea}.
\bjtitle{Mathematics and Mechanics of Solids}
\bvolume{28}(\bissue{1}),
\bfpage{196}--\blpage{207}
(\byear{2023})
\end{barticle}
\endbibitem

\bibitem[\protect\citeauthoryear{De~Bellis et~al.}{2023}]{debellis:2023}
\begin{botherref}
\oauthor{\bsnm{De~Bellis}, \binits{M.L.}},
\oauthor{\bsnm{Vasta}, \binits{M.}},
\oauthor{\bsnm{Gizzi}, \binits{A.}},
\oauthor{\bsnm{Pandolfi}, \binits{A.}}:
A numerical model of the human cornea accounting for the fiber-distributed
  collagen microstructure.
Mathematics and Mechanics of Solids,
10812865231202024
(2023)
\end{botherref}
\endbibitem

\bibitem[\protect\citeauthoryear{Pandolfi and De~Bellis}{2024}]{pandolfi:2024}
\begin{barticle}
\bauthor{\bsnm{Pandolfi}, \binits{A.}},
\bauthor{\bsnm{De~Bellis}, \binits{M.L.}}:
\batitle{Continuum versus micromechanical modeling of corneal biomechanics}.
\bjtitle{Mechanics of Materials}
\bvolume{199},
\bfpage{105162}
(\byear{2024})
\end{barticle}
\endbibitem

\bibitem[\protect\citeauthoryear{Simonini and Pandolfi}{2015}]{simonini:2015}
\begin{barticle}
\bauthor{\bsnm{Simonini}, \binits{I.}},
\bauthor{\bsnm{Pandolfi}, \binits{A.}}:
\batitle{Customized finite element modelling of the human cornea}.
\bjtitle{PloS one}
\bvolume{10}(\bissue{6}),
\bfpage{0130426}
(\byear{2015})
\end{barticle}
\endbibitem

\bibitem[\protect\citeauthoryear{Montanino et~al.}{2023}]{montanino:2023}
\begin{barticle}
\bauthor{\bsnm{Montanino}, \binits{A.}},
\bauthor{\bsnm{Overbeeke}, \binits{S.}},
\bauthor{\bsnm{Pandolfi}, \binits{A.}}:
\batitle{Modeling the biomechanics of laser corneal refractive surgery}.
\bjtitle{Journal of the Mechanical Behavior of Biomedical Materials}
\bvolume{145},
\bfpage{105998}
(\byear{2023})
\end{barticle}
\endbibitem

\bibitem[\protect\citeauthoryear{Holzapfel et~al.}{2000}]{holzapfel:2000}
\begin{barticle}
\bauthor{\bsnm{Holzapfel}, \binits{G.A.}},
\bauthor{\bsnm{Gasser}, \binits{T.C.}},
\bauthor{\bsnm{Ogden}, \binits{R.W.}}:
\batitle{A new constitutive framework for arterial wall mechanics and a
  comparative study of material models}.
\bjtitle{{Journal of Elasticity}}
\bvolume{61},
\bfpage{1}--\blpage{48}
(\byear{2000})
\end{barticle}
\endbibitem

\bibitem[\protect\citeauthoryear{Simo and Taylor}{1991}]{simo:1991}
\begin{barticle}
\bauthor{\bsnm{Simo}, \binits{J.C.}},
\bauthor{\bsnm{Taylor}, \binits{R.L.}}:
\batitle{Quasi-incompressible finite elasticity in principal stretches.
  continuum basis and numerical algorithms}.
\bjtitle{Computer Methods in Applied Mechanics and Engineering}
\bvolume{85}(\bissue{3}),
\bfpage{273}--\blpage{310}
(\byear{1991})
\doiurl{10.1016/0045-7825(91)90100-K}
\end{barticle}
\endbibitem

\bibitem[\protect\citeauthoryear{Rivlin and Saunders}{1951}]{rivlin:1951}
\begin{botherref}
\oauthor{\bsnm{Rivlin}, \binits{R.S.}},
\oauthor{\bsnm{Saunders}, \binits{D.W.}}:
Large elastic deformations of isotropic materials. {VII. E}xperiments on the
  deformation of rubber.
{Phylosophycal Transactions of The Royal Society of London, A}
\textbf{243}
(1951)
\end{botherref}
\endbibitem

\bibitem[\protect\citeauthoryear{K\"ory et~al.}{2024}]{koery:2024}
\begin{barticle}
\bauthor{\bsnm{K\"ory}, \binits{J.}},
\bauthor{\bsnm{Stewart}, \binits{P.S.}},
\bauthor{\bsnm{Hill}, \binits{N.A.}},
\bauthor{\bsnm{Luo}, \binits{X.-Y.}},
\bauthor{\bsnm{Pandolfi}, \binits{A.}}:
\batitle{A discrete-to-continuum model for the human cornea with application to
  keratoconus}.
\bjtitle{Royal Society Open Science}
\bvolume{11}(\bissue{7}),
\bfpage{240265}
(\byear{2024})
\end{barticle}
\endbibitem

\bibitem[\protect\citeauthoryear{Bonet and Wood}{2008}]{bonet:2008}
\begin{bbook}
\bauthor{\bsnm{Bonet}, \binits{J.}},
\bauthor{\bsnm{Wood}, \binits{R.D.}}:
\bbtitle{Nonlinear Continuum Mechanics for Finite Element Analysis},
\bedition{2}nd edn.
\bpublisher{Cambridge University Press},
\blocation{Cambridge}
(\byear{2008})
\end{bbook}
\endbibitem

\bibitem[\protect\citeauthoryear{Belytschko et~al.}{2014}]{belytschko:2014}
\begin{bbook}
\bauthor{\bsnm{Belytschko}, \binits{T.}},
\bauthor{\bsnm{Liu}, \binits{W.-K.}},
\bauthor{\bsnm{Moran}, \binits{B.}},
\bauthor{\bsnm{Elkhodary}, \binits{K.}}:
\bbtitle{Nonlinear Finite Elements for Continua and Structures}.
\bpublisher{John Wiley \& Sons},
\blocation{New Jersey}
(\byear{2014})
\end{bbook}
\endbibitem

\bibitem[\protect\citeauthoryear{Pandolfi and Holzapfel}{2008}]{pandolfi:2008}
\begin{barticle}
\bauthor{\bsnm{Pandolfi}, \binits{A.}},
\bauthor{\bsnm{Holzapfel}, \binits{G.A.}}:
\batitle{Three-dimensional modeling and computational analysis of the human
  cornea considering distributed collagen fibril orientations}.
\bjtitle{Journal of Biomechamical Engineering}
\bvolume{130},
\bfpage{061006}
(\byear{2008})
\end{barticle}
\endbibitem

\bibitem[\protect\citeauthoryear{Oakley and Knight~Jr}{1995}]{oakley:1995}
\begin{barticle}
\bauthor{\bsnm{Oakley}, \binits{D.R.}},
\bauthor{\bsnm{Knight~Jr}, \binits{N.F.}}:
\batitle{Adaptive dynamic relaxation algorithm for non-linear hyperelastic
  structures part i. formulation}.
\bjtitle{Computer methods in applied mechanics and engineering}
\bvolume{126}(\bissue{1-2}),
\bfpage{67}--\blpage{89}
(\byear{1995})
\end{barticle}
\endbibitem

\bibitem[\protect\citeauthoryear{Elsheikh}{2010}]{elsheikh:2010}
\begin{barticle}
\bauthor{\bsnm{Elsheikh}, \binits{A.}}:
\batitle{Finite element modeling of corneal biomechanical behavior}.
\bjtitle{{Journal of Refractive Surgery}}
\bvolume{26}(\bissue{4}),
\bfpage{289}--\blpage{300}
(\byear{2010})
\end{barticle}
\endbibitem

\bibitem[\protect\citeauthoryear{Ariza-Gracia et~al.}{2015}]{arizagracia:2015}
\begin{barticle}
\bauthor{\bsnm{Ariza-Gracia}, \binits{M.A.}},
\bauthor{\bsnm{Zurita}, \binits{J.F.}},
\bauthor{\bsnm{Pi\~nero}, \binits{D.P.}},
\bauthor{\bsnm{Rodriguez-Matas}, \binits{J.F.}},
\bauthor{\bsnm{Calvo}, \binits{B.}}:
\batitle{Coupled biomechanical response of the cornea assessed by non-contact
  tonometry. {A} simulation study}.
\bjtitle{{PLoS ONE}}
\bvolume{10}(\bissue{3}),
\bfpage{0121486}
(\byear{2015})
\end{barticle}
\endbibitem

\end{thebibliography}

\end{document}